\documentclass[11pt]{article}
\pdfoutput=1
\usepackage{graphicx}
\usepackage{amsmath,amssymb,amsthm,amsxtra,overpic,bbm,bm,epsfig}
\usepackage{color,subfigure}
\def\thefootnote{\fnsymbol{footnote}}

\begin{document}

\begin{flushright} 
IPPP/16/19  \\
\end{flushright} 

\vspace{0.2cm}

\begin{center}
{\Large\bf The role of flavon cross couplings in leptonic flavour mixing}
\end{center}

\vspace{0.2cm}

\begin{center}
{\bf Silvia Pascoli$^1$}\footnote{E-mail: silvia.pascoli@durham.ac.uk}
and 
{\bf Ye-Ling Zhou$^{1,2}$}\footnote{E-mail: ye-ling.zhou@durham.ac.uk}
\\
{$^1$ Institute for Particle Physics Phenomenology, Department of Physics, \\ Durham University, Durham DH1 3LE, United Kingdom} 
\\
{$^2$ Center for High Energy Physics, Peking University, Beijing 100080, China}
\end{center}

\vspace{1.5cm} 

\begin{abstract} 

In models with discrete flavour symmetries, flavons are critical to realise specific flavour structures. Leptonic flavour mixing originates from the misalignment of flavon vacuum expectation values which respect different residual symmetries in the charged lepton and neutrino sectors.  
Flavon cross couplings are usually forbidden, in order to protect these symmetries. Contrary to this approach, we show that cross couplings can play a key role and give raise to necessary corrections to flavour-mixing patterns, including a non-zero value for the reactor angle and CP violation. 
For definiteness, we present two models based on $A_4$. 
In the first model, all flavons are assumed to be real or pseudo-real, with 7 real degrees of freedom in the flavon sector in total. A sizable reactor angle associated with nearly maximal CP violation is achieved, and, as both originate from the same cross coupling, a sum rule results with a precise prediction for the value of the Dirac CP-violating phase. In the second model, the flavons are taken to be complex scalars, which can be connected with supersymmetric models and multi-Higgs models. The complexity properties of flavons provide new sources for generating the reactor angle. Models in this new approach introduce very few degrees of freedom beyond the Standard Model and can be more economical than those in the framework of extra dimension or supersymmetry. 

\end{abstract}

\begin{flushleft}
\hspace{0.8cm} PACS number(s): 14.60.Pq, 11.30.Hv, 12.60.Fr \\
\hspace{0.8cm} Keywords: Lepton flavour mixing, cross couplings, flavour symmetry
\end{flushleft}

\def\thefootnote{\arabic{footnote}}
\setcounter{footnote}{0}

\newpage

\section{Introduction}

Thanks to the discovery of neutrino oscillations \cite{atmospheric, solar}, leptonic flavour mixing has been observed by a series of atmospheric \cite{atmospheric}, solar \cite{solar}, accelerator \cite{accelerator} and reactor \cite{reactor} neutrino experiments. The three mixing angles have been measured to a very good accuracy. Both the atmospheric angle $\theta_{23}$ and the solar angle $\theta_{12}$ are rather large, and an order 0.1 reactor angle $\theta_{13}$ has been measured by reactor neutrino experiments \cite{PDG}. 
The $3\sigma$ ranges of mixing angles from current global analysis of solar, atmospheric, accelerator and reactor neutrino oscillation measurements \cite{globalfit} are given by 
\begin{eqnarray}
\sin^2\theta_{13} \in (0.0188,0.0251)\,,\quad
\sin^2\theta_{12} \in (0.270,0.344)\,,\quad
\sin^2\theta_{23} \in (0.385,0.644)\,.
\label{eq:3sigma}
\end{eqnarray}
There is also a preliminary hint \cite{globalfit, globalfit_former} for a maximally CP-violating value of the Dirac phase with a best-fit value $\delta\sim 270^\circ$ by combining the latest T2K \cite{T2K} and Daya Bay \cite{DayaBay} data, but the statistical significance of this result is still low and at $3\sigma$ all possible values of $\delta$ are allowed. 

Motivated by these values of the mixing angles, specific mixing patterns, realised at leading order, have been proposed in the last two decades. Among them, the tri-bimaximal (TBM) mixing predicts $\sin\theta_{12}=1/\sqrt{3}$ and $\sin\theta_{23}=1/\sqrt{2}$, which fit well current oscillation data and has therefore attracted a lot of attention \cite{TBM}. However, exact TBM mixing is ruled out due to the prediction of a vanishing $\theta_{13}$. To be compatible with current data, corrections to TBM must be introduced 
\begin{eqnarray}
\sin\theta_{13}=\sqrt{2}r'\,,\quad
\sin\theta_{12}=\frac{1}{\sqrt{3}}(1+s)\,,\quad
\sin\theta_{23}=\frac{1}{\sqrt{2}}(1+a)
\end{eqnarray}
with $r'\sim 0.1$ and $|s|, |a|\lesssim0.1$ \cite{King:2007pr}. Several possibilities have been discussed in the literature, such as TBM-Cabbibo mixing \cite{TBC} and trimaximal (TM) mixing in two forms, where the first and second columns of the PMNS matrix take the same values as TBM, called TM1 \cite{TM1} and TM2 \cite{TM2}, respectively. 

These specific mixing patterns can arise in flavour models in which discrete flavour symmetries are satisfied \cite{review}. These models suggest that, at some high energy scale, there exists an underlying discrete flavour symmetry in the flavour space. New scalars called flavons are introduced and achieve non-trivial vacuum expectation values (VEVs) at a lower energy scale, leading to the spontaneous breaking of the full symmetry. As proposed in \cite{Altarelli:2005yp}, \cite{Altarelli:2005yx} and later in \cite{Lam:2008sh}, in order to realise mixing patterns such as TBM, without imposing ad hoc relations among parameters, the flavour symmetry should be partly preserved. The residual or remnant symmetries in the charged lepton sector and the neutrino sector are different, they constrain the structures of the charged lepton and neutrino mass matrices, respectively, and eventually result in a specific leptonic mixing matrix. In order to obtain non-degenerate mass eigenvalues, these residual symmetries should be Abelian. They should be subgroups of the whole flavour symmetry at the high energy scale. In the charged lepton sector, the most economical choice is $Z_3$, and in the neutrino sector, the only choice is $Z_2$ or $Z_2\times Z_2'$,  if neutrinos are Majorana particles. To realise TBM, the generators for $Z_3$, $Z_2$ and $Z_2'$ are respectively given by \footnote{The explicit expressions of generators are basis-dependent. Here, we show them in the Altarelli-Feruglio basis, which can be found in \cite{Altarelli:2005yx}.}
\begin{eqnarray}
T=\left(
\begin{array}{ccc}
 1 & 0 & 0 \\
 0 & \omega ^2 & 0 \\
 0 & 0 & \omega  \\
\end{array}
\right)\,, \quad
S=\frac{1}{3} \left(
\begin{array}{ccc}
 -1 & 2 & 2 \\
 2 & -1 & 2 \\
 2 & 2 & -1 \\
\end{array}
\right) \,,\quad
U=\left(
\begin{array}{ccc}
 1 & 0 & 0 \\
 0 & 0 & 1 \\
 0 & 1 & 0 \\
\end{array}
\right)\,,
\label{eq:generators}
\end{eqnarray}
where $\omega=e^{i 2\pi/3}$. 
The most popular and simplest group to realise TBM is the tetrahedral group $A_4$, which is generated by $T$ and $S$.  The residual symmetries $Z_3$ generated by $T$ and $Z_2$ generated by $S$ are preserved in charged lepton sector and neutrino sector after $A_4$ breaking, respectively, while the other $Z_2'$ generated by $U$ arises in the neutrino sector accidentally. Many studies have been conducted on how to realise TBM and gain suitable corrections compatible with current oscillation data, for instance, see \cite{A4_TBM, TeV_flavour}. 

The flavon fields play a key role in the flavour model construction. To realise different residual symmetries in the charged lepton and neutrino sectors, we need different flavons, $\varphi$ and $\phi$, and require their VEVs be invariant under the action of $T$, $S$, respectively, i.e.,
\begin{eqnarray}
T\langle \varphi \rangle = \langle \varphi \rangle \,,\quad 
S\langle \phi \rangle = \langle \phi \rangle \,. 
\end{eqnarray}
The only solution of the above equation takes the following form
\begin{eqnarray}
\langle \varphi \rangle = \begin{pmatrix}1 \\ 0 \\ 0 \end{pmatrix} v_\varphi \,,\quad
\langle \phi \rangle = \begin{pmatrix} 1 \\ 1 \\ 1 \end{pmatrix} \frac{v_\phi}{\sqrt{3}} \,.
\label{eq:vev}
\end{eqnarray}
The flavon VEVs are dictated by the minimisation of the potential. The latter will generally contain cross couplings between $\varphi$ and $\phi$, which, although allowed by the full symmetry, would violate the residual ones. This vacuum alignment problem is a general problem of most flavour symmetry groups, not just limited to $A_4$ models.
Extra dimension or supersymmetry is invoked to forbid these cross couplings. In models with extra dimensions, $\varphi$ and $\phi$ can be localised on different branes such that they do not significantly couple with each other \cite{Altarelli:2005yp}. In models with supersymmetry, a continuous $U(1)_R$ symmetry and neutral scalars called driving fields are introduced \cite{Altarelli:2005yx}. The latter take a nontrivial charge under $U(1)_R$ and appear linearly in the superpotential. 
The minimisation of the flavon potential is finally simplified to vanishing $F$-terms of the driving fields. 
These approaches can solve the flavon VEV alignment problem effectively, but the price is that many degrees of freedom have to be introduced into the model. 
There is another solution by extending the flavour group, $H$, to a larger group $N\rtimes H$ \cite{large_group}. Here, $N\rtimes H$ should admit irreducible representations of $H$ such that the Standard Model leptons and one flavon can still transform in $H$,  while the other flavon transforms as a different representation that belongs to $N\rtimes H$ but not to $H$. 

Thanks to the flavon VEVs and the preservation of the two discrete symmetries in the charged lepton and neutrino sectors, TBM can be generated. However, due to the measurement of the reactor angle, the residual symmetries should be broken and corrections to TBM are needed in order to render models compatible with experimental data. 
In most models in the literature, this is realised by introducing higher dimensional operators, which may appear in both the flavon potential (especially the superpotential of flavons and driving fields in supersymmetric models) and couplings between leptons and flavons \cite{review}. 
These operators involve a certain scale $\Lambda$ higher than the scale of flavour symmetry characterised by $v_\varphi$ and $v_\phi$. They give rise to corrections to the mixing angles, e.g., $r'\sim v_\varphi/\Lambda$, implying that the new physics scale $\Lambda$ should not be far above the scale of flavour symmetries. 

In this paper, we will develop a new approach. 
Differently from above where the cross couplings are forbidden, we will allow their existence and investigate how they break the $Z_3$ and $Z_2$ residual symmetries. 
Similar ideas have been mentioned in Ref. \cite{TeV_flavour}, but a detailed discussion of how the vacuum is corrected by these terms and how the flavour mixing is affected is lacking. For definiteness, our flavour symmetry is assumed to be $A_4$. To correct TBM in agreement with the experimentally allowed region, cross couplings should be small, of order $\mathcal{O}(0.1)$. In this case, our calculation can be carried out perturbatively. Analytic relations between corrections of VEVs and the flavon VEV ratio $v_\varphi^2/ v_\phi^2$ will be derived. As a consequence, corrections of mixing angles to those in TBM are characterised by $v_\varphi^2/ v_\phi^2$ and the cross-coupling coefficients. 

The rest of this paper is organised as follows. In section \ref{sec:2}, we discuss the relation between the flavon potential and flavon VEVs in $A_4$. We first point out how the size of coefficients in the flavon potential determines the $Z_3$- and $Z_2$-symmetric VEVs, and then derive the corrections to these VEVs from cross couplings. Schematically, we present two flavour models in section \ref{sec:3}. In Model I, we introduce only two $A_4$ pseudo-real triplet flavons $\varphi, \phi$, one $A_4$ singlet flavon $\eta$ and an $A_4$ triplet right-handed neutrino $N$. This model is very economical since only 7 real degrees of freedom in the flavon sector are introduced. In Model II, we extend the flavons to complex fields and see how the mixing structure deviates from the pseudo-real flavon case. This extension is interesting since, if we wanted to draw a connection with supersymmetric and multi-Higgs models, flavons must be complex. We summarise the results in section \ref{sec:4}. In appendix \ref{sec:A}, we list the full solutions of VEVs for a single $A_4$ triplet flavon. 

\section{Flavon cross couplings and vacuum alignment \label{sec:2}}

We assume the flavour symmetry to be the tetrahedral group $A_4$ \cite{Ma:2001dn}, the group of even permutations of four objects. It is generated by $S$ and $T$ with the requirement $S^2=T^3=(ST)^3=1$, and contains 12 elements: $T$, $ST$, $TS$, $STS$, $T^2$, $ST^2$, $T^2S$, $TST$, $S$, $T^2ST$, $TST^2$ and the identity element $1$. It is the smallest discrete group which has a 3-dimensional (3d) irreducible representation $\mathbf{3}$ and in this representation, the generators $S$ and $T$ can be given as in Eq. \eqref{eq:generators}, respectively. Besides, it has three 1-dimensional irreducible representations: the trivial singlet $\mathbf{1}$ and non-trivial singlets $\mathbf{1'}$, $\mathbf{1}''$. The Kronecker product of two 3d irreducible representations can be reduced: $\mathbf{3}\times\mathbf{3}=\mathbf{1}+\mathbf{1}'+\mathbf{1}''+\mathbf{3}_S+\mathbf{3}_A$, where the subscripts $_S$ and $_A$ stand for the symmetric and anti-symmetric parts, respectively. 

We introduce a flavon field $\varphi=(\varphi_1, \varphi_2, \varphi_3)^T$. It contains three gauge-singlet scalars and  transforms as a pseudo-real triplet $\mathbf{3}$ representation of $A_4$ which requires $\varphi_1^*=\varphi_1$ and $\varphi_2^*=\varphi_3$.  
The renormalisable flavon potential invariant under $A_4$ is generically written as
\begin{eqnarray}
V(\varphi)= \frac{1}{2}\mu^2_\varphi (\varphi \varphi)_\mathbf{1} +\frac{1}{4} \left[f_1 \big( (\varphi \varphi)_\mathbf{1}\big)^2 + f_2 (\varphi \varphi)_{\mathbf{1}'} (\varphi \varphi)_{\mathbf{1}''} + f_3 \big( (\varphi \varphi)_{\mathbf{3}_S} (\varphi \varphi)_{\mathbf{3}_S} \big)_\mathbf{1} \right]\,,
\label{eq:Vvarphi}
\end{eqnarray}
where all the coefficients $\mu_\varphi^2$ and $f_{1,2,3}$ are real. The conditions $\mu_\varphi^2<0$, $f_1+f_2>0$ and $f_1+f_3>0$ are required to achieve a nontrivial and stable vacuum. 
For the notation of representations and detailed expression of $V(\varphi)$, please see Eq. \eqref{eq:CG2} in appendix \ref{sec:A}.
To simplify our discussion, tri-linear terms such as $\big( (\varphi \varphi)_{\mathbf{3}_S} \varphi \big)_\mathbf{1}$ are not considered here. These terms can be forbidden by an additional $Z_2$ symmetry ($\varphi\to-\varphi$). 
There are two classes of configurations that are candidates for the vacuum of $\varphi$. They are characterised by
\begin{eqnarray}
\langle \varphi \rangle_1 = \begin{pmatrix}1 \\ 0 \\ 0 \end{pmatrix} v_{\varphi1} \,,\qquad
\langle \varphi \rangle_2 = \begin{pmatrix} 1 \\ 1 \\ 1\end{pmatrix} \frac{v_{\varphi2}}{\sqrt{3}} \,,  
\label{eq:vevs1}
\end{eqnarray}
where 
\begin{eqnarray}
v_{\varphi1}^2= \frac{-\mu_\varphi^2}{f_1+f_3}\,, &\qquad&
v_{\varphi2}^2= \frac{-\mu_\varphi^2}{f_1+f_2} \,.
\label{eq:vevs2}
\end{eqnarray}
The potential $V(\varphi)$ takes extremal values
\begin{eqnarray}
V(\langle \varphi \rangle_1)=-\frac{\mu_\varphi^4}{4 (f_1+f_3)}\,, &\qquad&
V(\langle \varphi \rangle_2)=-\frac{\mu_\varphi^4}{4 (f_1+f_2)} \,,
\end{eqnarray}
at $\langle \varphi \rangle_1$ and $\langle \varphi \rangle_2$, 
respectively, which depend on the relative size of $f_2$ and $f_3$. In the case $f_2>f_3$, $V(\varphi)$ has the global minimum value at $\langle \varphi \rangle_1$, and $\langle \varphi \rangle_2$ is just an unstable saddle point. Thus, the vacuum of $\varphi$ is $\langle \varphi \rangle_1$. In the opposite case $f_2<f_3$, $\langle \varphi \rangle_2$ is the vacuum. 
For a detailed discussion of the determination of the vacuum of the potential $V(\varphi)$, see appendix \ref{sec:A}. From here onward, we assume $f_2>f_3$ such that the VEV of $\varphi$ is fixed at $\langle \varphi \rangle_1$ at leading order. 

Then, we consider another $A_4$ pseudo-real triplet scalar $\phi=(\phi_1,\phi_2,\phi_3)^T$. Its potential $V(\phi)$ takes the same form as $V(\varphi)$ with coefficients $\mu_\varphi$ and $f_i$ replaced by $\mu_\phi$ and $g_i$, respectively. All the results above for $\varphi$ will apply with the substitutions $\varphi\to \phi$ and $f_i \to g_i$. In order to select the other vacuum alignment, we assume $g_2<g_3$, so that $V(\phi)$ takes the global minimum value at $\langle \phi \rangle_2$. 

With the assumptions $f_2> f_3$ and $g_2<g_3$, we obtain the VEVs of $\varphi$ and $\phi$ at $\langle \varphi \rangle_1$ and $\langle \phi \rangle_2$, respectively, as in Eq. \eqref{eq:vev}. They respect the $Z_3$ and $Z_2$ residual symmetries, respectively, and reproduce the TBM mixing pattern. 

Cross couplings between $\varphi$ and $\phi$ would modify the VEVs of $\varphi$ and $\phi$. The most general Lagrangian describing flavon cross couplings is given by
\begin{eqnarray}
V(\varphi, \phi) &=& \frac{1}{2} \epsilon_1( \varphi \varphi )_\mathbf{1} ( \phi \phi )_\mathbf{1} + \frac{1}{4}\big[ \epsilon_2 ( \varphi \varphi )_{\mathbf{1}''} ( \phi \phi )_{\mathbf{1}'} +\text{h.c.} \big]
+ \frac{1}{2} \epsilon_3 \big( ( \varphi \varphi )_{\mathbf{3}_S} (\phi \phi )_{\mathbf{3}_S} \big)_{\mathbf{1}} \,,
\label{eq:Vmix} 
\end{eqnarray}
where $\epsilon_{1,3}$ are real and $\epsilon_2$ is the only complex parameter in the flavon potential. 
By including these couplings, we get the full renormalisable flavon potential invariant under the flavour symmetry $A_4\times Z_2$, i.e., $V=V(\varphi)+V(\phi)+V(\varphi,\phi)$. 
The cross couplings will break the residual symmetries. To achieve order $\mathcal{O}(0.1)$ corrections to the TBM mixing, we assume the coefficients $\epsilon_i$ to be of the same order. In this case, modifications of the flavon VEVs are small and the residual $Z_2$ and $Z_3$ symmetries are preserved at leading order. 

As the cross couplings are assumed to be small, we can proceed to compute analytically the corrections to the leading terms. Expanding the VEVs as 
\begin{eqnarray}
\langle \varphi_1 \rangle = v_\varphi+ \delta v_{\varphi1} \,, \hspace{5mm}
\langle \varphi_2 \rangle =  \delta v_{\varphi2} \,, \hspace{14mm}
\langle \varphi_3 \rangle =  \delta v_{\varphi3} \,, \hspace{10mm} \nonumber\\
\langle \phi_1 \rangle = \frac{v_\phi}{\sqrt{3}}+ \delta v_{\phi1} \,, \quad
\langle \phi_2 \rangle = \frac{v_\phi}{\sqrt{3}}+ \delta v_{\phi2} \,, \quad
\langle \phi_3 \rangle = \frac{v_\phi}{\sqrt{3}}+ \delta v_{\phi3} \,, 
\label{eq:expansion}
\end{eqnarray} 
we retain the quadratic terms of $V(\varphi)$ and $V(\phi)$ and the linear terms of $V(\varphi,\phi)$, which are the only ones relevant to the vacuum shifts at first order: 
\begin{eqnarray}
V^{(2)}(\varphi) &=& \frac{1}{2}m_{\varphi1}^2 \delta v_{\varphi1}^2 + m_{\varphi2}^2 |\delta v_{\varphi2}|^2  \,,\nonumber\\
V^{(2)}(\phi) &=& \frac{1}{6}(m_{\phi1}^2 + 2 m_{\phi2}^2 ) \left[\delta v_{\phi1}^2 + 2| \delta v_{\phi2}|^2 \right] + \frac{1}{6}(m_{\phi1}^2 - m_{\phi2}^2 ) \left(\delta v_{\phi2}^2 + 2 \delta v_{\phi1} \delta v_{\phi2}^*+ \text{h.c.}\right)  \,,\nonumber\\
V^{(1)}(\varphi, \phi) &=& \epsilon_1 v_\phi^2 v_\varphi \delta v_{\varphi1} + \epsilon_1 v_\varphi^2  \frac{v_\phi}{\sqrt{3}} ( \delta v_{\phi1}+\delta v_{\phi2}+ \delta v_{\phi2}^* ) \nonumber\\
&&+ \frac{1}{2} v_\phi^2v_\varphi( \epsilon_2 \delta v_{\varphi2}^*+ \epsilon_2^* \delta v_{\varphi2})
+ \frac{1}{2} \epsilon_3  v_\varphi^2 \frac{v_\phi}{\sqrt{3}} ( 2 \delta v_{\phi1} - \delta v_{\phi2} - \delta v_{\phi2}^* ) \,,
\label{eq:Vmix_quadratic}
\end{eqnarray}
where
\begin{eqnarray}
m_{\varphi1}^2 = 2(f_1+f_3)v_\varphi^2 \,, &&
m_{\varphi2}^2 = (f_2-f_3) v_\varphi^2\,, \nonumber\\
m_{\phi1}^2 = 2(g_1+g_2)v_\phi^2 \,, &&
m_{\phi2}^2 = \frac{3}{2}(g_3-g_2) v_\phi^2\,.
\label{eq:flavon_mass1}
\end{eqnarray} 
One can check that $\varphi_1$ and $\varphi_2$ are actually the mass eigenstates of $\varphi$ after $A_4$ breaking to $Z_3$, with mass eigenvalues $m_{\varphi1}^2$ and $m_{\varphi2}^2$, respectively. However, $\phi_1$ and $\phi_2$ are not mass eigenstates of $\phi$. By diagonalising the mass matrix of $\phi$, we derive the mass eigenvalues $m_{\phi1}^2$ and $m_{\phi2}^2$. 
We minimise the potential in Eq. \eqref{eq:Vmix_quadratic} and derive the modified VEVs of $\varphi$ and $\phi$ to be
\begin{eqnarray}
\langle \varphi \rangle \approx \begin{pmatrix}1 \\ \epsilon_{\varphi} \\ \epsilon_{\varphi}^* \end{pmatrix} v_\varphi \,,\quad
\langle \phi \rangle \approx \begin{pmatrix} 1-2 \epsilon_\phi \\ 1+\epsilon_\phi \\ 1+\epsilon_\phi\end{pmatrix} \frac{v_\phi}{\sqrt{3}} \,, \quad 
\label{eq:flavon_vevs}
\end{eqnarray}
where
\begin{eqnarray}
&& v_{\varphi}^2= \frac{-\mu_\varphi^2}{f_1+f_3}\,, \qquad
v_{\phi}^2= \frac{-\mu_\phi^2}{g_1+g_2}\,, \nonumber\\
&& \epsilon_\varphi= - \frac{ \epsilon_2 }{2 } \frac{v_\phi^2}{m_{\varphi_2}^2} \,, \;\;\quad
\epsilon_\phi = \frac{ \epsilon_3}{2} \frac{v_\varphi^2}{m_{\phi_2}^2} \,. 
\label{eq:vevs&epsilon}
\end{eqnarray}
Here, we have redefined the effective $\mu_\varphi^2$ and $\mu_\phi^2$ via $\mu_\varphi^2+\epsilon_1 v_\phi^2 \to \mu_\varphi^2$ and $\mu_\phi^2+\epsilon_1 v_\varphi^2 \to \mu_\phi^2$ to absorb $\epsilon_1$, since the $\epsilon_1$ term is a trivial correction keeping the residual symmetries unchanged no matter what kind of VEVs $\varphi$ and $\phi$ achieve. 
The $\epsilon_2$ term is the main source of the breaking of the $Z_3$ symmetry in $\langle \varphi \rangle$, since $\langle (\phi\phi)_{\mathbf{1}''} \rangle$ in this term approximates to $v_\phi^2$, which is not invariant under the action of $T$. $\langle (\phi\phi)_{\mathbf{3}_S} \rangle$ in the $\epsilon_3$ term vanishes and will not break the $Z_3$ symmetry at leading order. The $\epsilon_3$ term is the main source of breaking $Z_2$ in $\langle \phi \rangle$. Similarly, the reason is that $\langle (\varphi\varphi)_{\mathbf{3}_S} \rangle\approx(1,0,0)^T v_\varphi^2/3$ in this term is not invariant under the action of $S$, while $\langle (\varphi\varphi)_{\mathbf{1}''} \rangle$ in the $\epsilon_2$ term vanishes at leading order. 

The effective complex parameter $\epsilon_\varphi$, which measures the amount of $Z_3$ breaking, is a crucial parameter in our discussion. We parameterize it as $\epsilon_\varphi=|\epsilon_\varphi|e^{i\theta_\varphi}$, in which $-180^\circ<\theta_{\varphi}\leqslant180^\circ$. In the next section, we will construct two lepton flavour models with lepton mixing parameters corrected by $\epsilon_\varphi$.   

\section{Flavour Models \label{sec:3}} 

In this section, we will construct two leptonic flavour models, introducing gauge singlets $N$ and implementing the seesaw mechanism. In both models, we assume the flavour symmetry to be $A_4\times Z_2\times Z_4$. The additional $Z_4$ is imposed to forbid unnecessary couplings between flavons and leptons. We introduce only two $A_4$ triplet flavons $\varphi$ and $\phi$ and one singlet flavon $\eta$. The singlet $\eta$ is used to give suitable neutrino mass spectra. We will exploit the presence of cross couplings between $\varphi$ and $\phi$ to realise lepton flavour mixing compatible with the constraints of neutrino oscillation experiments. The main difference of these two models is that $\varphi$ and $\phi$ in Model I transform as pseudo-real triplets of $A_4$, and those in Model II are complex triplets of $A_4$. Although the trivial singlet flavon $\eta$ is real in Model I and complex in Model II, it does not result in any different mixing structure in the two models. 

\subsection{Model I} 

In model I, both $\varphi$ and $\phi$ are pseudo-real, and $\eta$ is a real singlet. This model is very economical as only 7 real degrees of freedom in the flavon sector are introduced. 
Transformation properties of $\varphi$, $\phi$ and $\eta$ under $A_4\times Z_2\times Z_4$, together with those of the Higgs $H$, leptons in the Standard Model $\ell_L$, $e_R$, $\mu_R$, $\tau_R$ and an extra right-handed neutrino $N$ are shown in Table \ref{tab:fields}. 


\begin{table}[h]
\begin{center}
\begin{tabular}{c c c c c c c c c c}
\hline\hline
Fields & $\ell_L$ & $e_R,\mu_R,\tau_R$ & $N$ & $H$ & $\varphi$
& $\phi$ & $\eta$    \\ \hline

$A_4$ & $\mathbf{3}$ & $\mathbf{1},\mathbf{1}',\mathbf{1}''$ & $\mathbf{3}$ & $\mathbf{1}$ & $\mathbf{3}$ & $\mathbf{3}$ & $\mathbf{1}$  \\

$Z_2$ & $1$ & $-1$ & $1$ & $1$ & $-1$ & $1$ & $1$  \\

$Z_4$ & $i$ & $i$ & $i$ & $1$ & $1$ & $-1$ & $-1$ \\

\hline
\hline
\end{tabular}
\caption{\label{tab:fields} Representations of the fields under the flavour symmetry $A_4 \times Z_2 \times Z_4$.}
\end{center}
\end{table}

As discussed in the last section, the flavon fields $\varphi$ and $\phi$ get VEVs as shown in Eq. \eqref{eq:flavon_vevs}. Cross couplings of $\varphi$, $\phi$ with $H$ and the new flavon singlet $\eta$ will modify the VEVs of $\varphi$ and $\phi$. Since $H$ and $\eta$ are flavour singlets $\mathbf{1}$, their cross couplings with $\varphi$ and $\phi$ can only modify the overall sizes of $\langle \varphi \rangle$ and $\langle \phi \rangle$, i.e., $v_\varphi$ and $v_\phi$, but have no influence on the direction $(1, \epsilon_\varphi, \epsilon_\varphi^*)^T$ and $(1-2\epsilon_\phi, 1+\epsilon_\phi,1+\epsilon_\phi)^T$. 
The effect of $H$ and $\eta$ can be reabsorbed in the redefinition of $\mu_\varphi^2$ and $\mu_\phi^2$ and, having done so, the expressions of $v_\varphi$, $v_\phi$, and $\epsilon_\varphi$, $\epsilon_\phi$ in Eq. \eqref{eq:vevs&epsilon} remain valid.

The Lagrangian terms for generating lepton masses are given by
\begin{eqnarray}
-\mathcal{L}_l &=& \frac{y_e}{\Lambda} (\overline{\ell_L} \varphi)_\mathbf{1} e_R H + \frac{y_\mu}{\Lambda} (\overline{\ell_L} \varphi)_{\mathbf{1}''} \mu_R H + \frac{y_\tau}{\Lambda} (\overline{\ell_L} \varphi)_{\mathbf{1}'} \tau_R H + \text{h.c.} + \cdots \,, \nonumber\\
-\mathcal{L}_\nu &=& y_D (\overline{\ell_L}N)_\mathbf{1} \tilde{H}  + \frac{y_1}{2} \big((\overline{N^c} N)_{\mathbf{3}_S} \phi \big)_\mathbf{1} + \frac{y_2}{2} (\overline{N^c} N)_\mathbf{1} \eta + \text{h.c.} + \cdots \,,
\label{eq:Lagrangian}
\end{eqnarray}
where the dots stand for higher dimensional operators. Note that  in many models, higher dimensional operators are important because they are responsible for corrections to leading order mixing structure. 
For example, the dimension-6 operator $\big((\overline{N^c} N)_{\mathbf{3}_S} ((\varphi\varphi)_{\mathbf{3}_S} \phi)_{\mathbf{3}_S} \big)_\mathbf{1}/\Lambda^2$ will modify flavour mixing from the leading order structure since the vacuum direction of the combined term $\langle ((\varphi\varphi)_{\mathbf{3}_S} \phi)_{\mathbf{3}_S} \rangle \sim  (2,-1,-1)^T $ does not preserve the $Z_2$ symmetry. Compared with these models, our model assumes that higher dimensional operators are negligible such that they cannot lead to significant modifications.

Leptons gain masses with specific mass matrix structures, after breakings of the flavour symmetry and the electroweak symmetry. The resulting charged lepton mass matrix can be written as
\begin{eqnarray}
M_l \approx \left(
\begin{array}{ccc}
 y_e &  y_\mu \epsilon_\varphi^* &  y_\tau \epsilon_\varphi \\
 y_e \epsilon_\varphi &  y_\mu &  y_\tau \epsilon_\varphi^* \\
 y_e \epsilon_\varphi^* &  y_\mu \epsilon_\varphi &  y_\tau \\
\end{array}
\right)\frac{v v_\varphi}{\sqrt{2}\Lambda} \,,
\label{eq:lepton_mass}
\end{eqnarray}
where $\langle H \rangle=v/\sqrt{2}$ with $v=246$ GeV being the VEV of the Higgs boson. The Dirac and Majorana mass matrices for neutrinos are given by
\begin{eqnarray}
&&M_D=\left(
\begin{array}{ccc}
 1 & 0 & 0 \\
 0 & 1 & 0 \\
 0 & 0 & 1 \\
\end{array}
\right) y_D \frac{v}{\sqrt{2}}
\,,\nonumber\\
&&M_N \approx \frac{1}{\sqrt{3}}
\left(
\begin{array}{ccc} 
\sqrt{3}  y_2v_\eta+ y_1v_\phi (1-2 \epsilon_\phi) & - \frac{1}{2} y_1v_\phi (1+\epsilon_\phi) & - \frac{1}{2} y_1v_\phi (1+\epsilon_\phi) \\
 - \frac{1}{2} y_1v_\phi (1+\epsilon_\phi) & y_1v_\phi (1+\epsilon_\phi) & \sqrt{3} y_2v_\eta- \frac{1}{2} y_1v_\phi (1-2 \epsilon_\phi) \\
 - \frac{1}{2} y_1v_\phi (1+\epsilon_\phi) & \sqrt{3} y_2v_\eta- \frac{1}{2} y_1v_\phi (1-2 \epsilon_\phi) & y_1v_\phi (1+\epsilon_\phi) \\
\end{array}
\right)\,,
\label{eq:neutrino_mass}
\end{eqnarray}
where $v_\eta$ the VEV of the singlet $\eta$, and the active neutrinos obtain masses through the seesaw mechanism
\begin{eqnarray}
M_\nu=-M_D M_N^{-1}M_D^T\,.
\end{eqnarray}
After diagonalising $M_l$ and $M_\nu$, we compute the masses of the charged leptons and neutrinos
\begin{eqnarray}
&\displaystyle m_e \approx |y_e v_\varphi| \frac{v}{\sqrt{2}\Lambda}\,,\quad
m_\mu \approx |y_\mu v_\varphi| \frac{v}{\sqrt{2}\Lambda}\,,\quad
m_\tau \approx |y_\tau v_\varphi| \frac{v}{\sqrt{2}\Lambda}\,, \nonumber\\
&\displaystyle m_1 \approx \frac{2|y_D^2|v^2}{| \sqrt{3} y_1 v_\phi + 2 y_2 v_\eta |}\,, \quad 
m_2 \approx \frac{|y_D^2| v^2}{ |y_2 v_\eta|}\,, \quad 
m_3 \approx \frac{2|y_D^2|v^2}{| \sqrt{3} y_1 v_\phi - 2 y_2 v_\eta |} \,.
\end{eqnarray}
The PMNS matrix is obtained by the product of the unitary matrices which diagonalise the charged lepton and neutrino mass matrices and is given by
\begin{eqnarray}
U_\text{PMNS} \approx \left(
\begin{array}{ccc}
 1 & -\epsilon_\varphi^* & -\epsilon_\varphi \\
 \epsilon_\varphi & 1 & -\epsilon_\varphi^* \\
 \epsilon_\varphi^* & \epsilon_\varphi & 1 \\
\end{array}
\right)
\left(
\begin{array}{ccc}
 \frac{2}{\sqrt{6}} & \frac{1}{\sqrt{3}} & 0 \\
 -\frac{1}{\sqrt{6}} & \frac{1}{\sqrt{3}} & -\frac{1}{\sqrt{2}} \\
 -\frac{1}{\sqrt{6}} & \frac{1}{\sqrt{3}} & \frac{1}{\sqrt{2}} \\
\end{array}
\right)
\left(
\begin{array}{ccc}
 1 & \sqrt{2}\epsilon_\phi & 0 \\
 -\sqrt{2} \epsilon_\phi & 1 & 0 \\
 0 & 0 & 1 \\
\end{array}
\right) P_\nu \,.
\label{eq:PMNS}
\end{eqnarray} 
The LHS of the PMNS matrix is the correction from the charged lepton sector, the middle is the TBM mixing, and the RHS is the correction from the neutrino sector.
The mixing angles can be expressed in terms of the model parameters $\theta_\varphi$, $|\epsilon_\varphi|$ and $\epsilon_\phi$ as 
\begin{eqnarray} 
&&\sin\theta_{13} \approx \sqrt{2}|\epsilon_\varphi\sin\theta_{\varphi}|\,,\nonumber\\
&&\sin\theta_{12} \approx \frac{1}{\sqrt{3}} \big( 1-2|\epsilon_\varphi|\cos\theta_\varphi+2\epsilon_\phi \big) \,,\nonumber\\
&&\sin\theta_{23} \approx \frac{1}{\sqrt{2}} \big( 1+|\epsilon_\varphi|\cos\theta_\varphi \big) \,,
\label{eq:mixing_angles}
\end{eqnarray} 
and the CP-violating phases are approximately given by
\begin{eqnarray}
&&\delta \approx \left\{\begin{array}{c} 270^\circ -2 |\epsilon_\varphi| \sin\theta_\varphi\,, \quad \theta_\varphi>0 \,, \\ \;\;90^\circ -2 |\epsilon_\varphi| \sin\theta_\varphi\,, \quad \theta_\varphi<0 \,,
 \end{array} \right. \nonumber\\
&& \alpha_{21} \approx \text{Arg}\left\{1+\frac{\sqrt{3} y_1 v_\phi}{ 2 y_2 v_\eta} \right\} \,, \nonumber\\
&& \alpha_{31} \approx \text{Arg}\left\{ \Big[\frac{\sqrt{3} y_1 v_\phi + 2 y_2 v_\eta}{\sqrt{3} y_1 v_\phi - 2 y_2 v_\eta} \Big] \Big[ 1- 4 i |\epsilon_\varphi| \sin\theta_\varphi \Big]  \right\}\,,
\label{eq:CP_phases}
\end{eqnarray}
where the Majorana phases $\alpha_{21}$ and $\alpha_{31}$ are defined in Ref. \cite{PDG}. If there are no cross couplings between $\varphi$ and $\phi$, we obtain an explicit TBM mixing, which predicts $\sin\theta_{13}=0$, $\sin\theta_{12}=1/\sqrt{3}$ and $\sin\theta_{23}=1/\sqrt{2}$. Differently from most models in the literature in which corrections are functions of ratio of the flavon VEV to some unknown high energy scale, e.g., $v_\varphi/\Lambda$, the corrections here are functions of the ratio of the two flavon VEVs $v_\phi/v_\varphi$, as well as the coefficients $\epsilon_2$ and $\epsilon_3$. 

\begin{figure}[h!]
\begin{center}
\includegraphics[width=1\textwidth,natwidth=600,natheight=600]{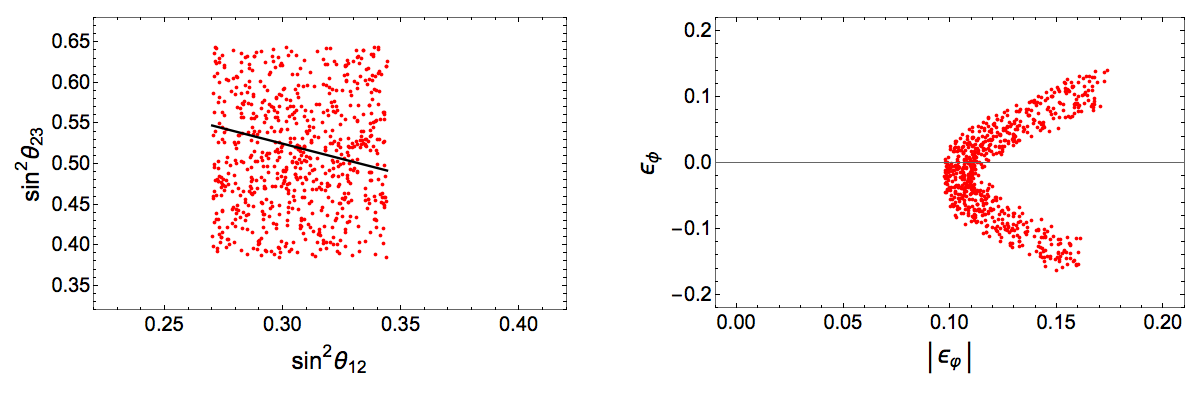}
\vspace{-1cm} 
\caption{Theoretical prediction of mixing angles (left panel) and the allowed parameter space of $|\epsilon_\varphi|$ and $\epsilon_\phi$ (right panel) in Model I. The $3\sigma$ range data of mixing angles in Ref. \cite{globalfit} have been used as the cut. The straight line in the left panel stands for the correlation of $\theta_{12}$ vs $\theta_{23}$ in the limit $\epsilon_\phi=0$ with the expression given in Eq. \eqref{eq:corr_mixing}. }
\label{fig:mixI}
\end{center}
\end{figure}

We can also express the corrections in terms of the parameters $r^\prime$, $s$ and $a$ introduced earlier as
\begin{eqnarray}
r'=|\epsilon_\varphi\sin\theta_{\varphi}| \,, \quad
s=-2|\epsilon_\varphi|\cos\theta_\varphi+2\epsilon_\phi \,, \quad
a=|\epsilon_\varphi|\cos\theta_\varphi \,.
\end{eqnarray} 
A sum rule for the corrections to $\theta_{13}$ and $\theta_{23}$ 
\begin{eqnarray}
r^{\prime2}+a^2=|\epsilon_\varphi|^2\,
\label{eq:sumrule_ra}
\end{eqnarray}  
is obtained. From this relation in Eq. \eqref{eq:sumrule_ra}, we can deduce that the value of $|\epsilon_\varphi|$ is around 0.1-0.2, consistently with the initial assumptions made on this parameter. The non-zero reactor angle $\theta_{13}$ arises from the imaginary part of $\epsilon_\varphi$, i.e., the cross coupling $\text{Im}(\epsilon_2) ( \varphi \varphi )_{\mathbf{1}''} ( \phi \phi )_{\mathbf{1}'}$. 
To be compatible with the measured value of $\sin^2\theta_{13}\approx0.02$ \cite{globalfit}, the effective parameter $|\epsilon_\varphi|\sin\theta_\varphi$ should be around $\pm0.1$, implying that $\sin \theta_\varphi$ cannot be too small. The other two parameters $|\epsilon_\varphi|\cos\theta_\varphi$ and $\epsilon_\phi$ can take a value $\lesssim0.1$, depending on the precise measured values of $\theta_{12}$ and $\theta_{23}$. We show the numerical results for the correlation between $\theta_{12}$ and $\theta_{23}$ and the allowed parameter space of $|\epsilon_\varphi|$ vs $\epsilon_\phi$ of Model I in  Fig. \ref{fig:mixI}. Here, $|\epsilon_{\varphi}|$, $\vartheta_\varphi$, and $\epsilon_{\phi}$ are randomly generated in the range $[0,\,0.2]$, $[0,\,360^\circ)$ and $[-0.2,\,0.2]$, respectively, and those compatible with the $3\sigma$ range data of the mixing angles in Eq. \eqref{eq:3sigma} are shown in the figure. We see that $0.1 \lesssim |\epsilon_\varphi| \lesssim 0.17$ and $-0.18 \lesssim \epsilon_\phi \lesssim 0.17$ are required. In the region where $\epsilon_\phi$ has a large deviation from 0, the correlation $|\epsilon_\phi|\propto |\epsilon_\varphi|$ roughly holds, which is required by the constraint of $\theta_{12}$, as shown in Eq. \eqref{eq:mixing_angles}. The phase of $\epsilon_\varphi$ is given by
\begin{eqnarray}
|\tan\theta_\varphi| = \frac{r'}{|a|} = \frac{\sin\theta_{13}}{|2\sin\theta_{23}-\sqrt{2}|} \,.
\end{eqnarray}
Taking the $3\sigma$ ranges of $\theta_{13}$ and $\theta_{23}$ in Eq. \eqref{eq:3sigma} into account, we get $38^\circ<|\theta_\varphi|<142^\circ$. 

The Dirac CP-violating phase has a small deviation from the value relative to maximal CP violation: 
\begin{eqnarray}
&&\delta \approx \left\{\begin{array}{c} 270^\circ - \sqrt{2} \theta_{13}\,, \quad \theta_\varphi>0 \,, \\ \;\;90^\circ +\sqrt{2}\theta_{13}\,, \quad \theta_\varphi<0 \,. 
 \end{array} \right. 
\end{eqnarray}
The connection between $\delta$ and $\theta_{13}$ can be expected since they both result from the same cross coupling $\text{Im}(\epsilon_2)(\varphi\varphi)_{\mathbf{1}''}(\phi\phi)_{\mathbf{1}'}$. 
Taking the best-fit value of $\theta_{13}$, we predict $\delta=258^\circ$ or $102^\circ$, which is very close to the current best-fit value in the inverted mass ordering. The Majorana CP-violating phases take values totally independent of the other mixing parameters. They can be arbitrary, depending on the relative phase of $y_1/y_2$. 

This model can be further simplified in the limit $\epsilon_\phi\to0$. Due to the fact that the ratio $\epsilon_\varphi/\epsilon_\phi$ is proportional to $v_\phi^4/v_\varphi^4$, a small hierarchy between $v_\phi$ and $v_\varphi$ would result in a large hierarchy between $\epsilon_\varphi$ and $\epsilon_\phi$. Therefore, if $v_\phi$ is significantly larger than $v_\varphi$, $\epsilon_\phi$ will be much smaller than $\epsilon_\varphi$, and its contribution to the mixing angles can be neglected. In this case, all deviations from the tri-bimaximal mixing pattern come from the corrections from the charged lepton sector. A simple relation between $\theta_{12}$ and $\theta_{23}$ is derived
\begin{eqnarray}
3\sin^2\theta_{12} + 4\sin^2\theta_{23} =3\,,
\label{eq:corr_mixing}
\end{eqnarray} 
or equivalently, $s+2a=0$. 
Taking the $3\sigma$ range of $\theta_{12}$ to the sum rule in Eq. \eqref{eq:corr_mixing} predicts $0.492<\sin^2\theta_{23}<0.548$, the second octant of $\theta_{23}$ being preferred. Finally, let us discuss the constraint on $\epsilon_\varphi$ in this simplified case. $|\epsilon_\varphi|$ must be in general in the range from 0.1 to 0.12, as shown in Fig. \ref{fig:mixI}. The phase of $\epsilon_\varphi$ can be further constrained by the mixing angle $\theta_{12}$: 
\begin{eqnarray}
|\tan\theta_{\varphi}|=\frac{2r'}{|s|} = \frac{\sqrt{2}\sin\theta_{13}}{|\sqrt{3}\sin\theta_{12}-1|}\,.
\end{eqnarray}
Using the $3\sigma$ ranges of $\theta_{12}$ and $\theta_{13}$, $\theta_\varphi$ is constrained to be $63^\circ<|\theta_\varphi|<117^\circ$. This leads to $|\sin\theta_\varphi|\approx 1$. Therefore,  
\begin{eqnarray}
\sin\theta_{13}\approx \sqrt{2} |\epsilon_\varphi|
\end{eqnarray}
with $|\epsilon_\varphi| \approx 0.1$ roughly holding. Specifically, taking the current experimental best-fit values $\sin^2\theta_{12}=0.304$ and $\sin^2\theta_{13}=0.0219$ \cite{globalfit}, we predict $\sin^2 \theta_{23} = 0.52$ in the second octant, with $\theta_\varphi \simeq \pm 78^\circ$ and $|\epsilon_\varphi| \simeq 0.106$. 

\subsection{Model II} 

In Model I, we have used flavon cross couplings to obtain flavour mixing deviations from the TBM mixing. Connections between corrections of mixing angles and the ratio of flavon VEVs, i.e., $\delta\theta_{ij}=f(v_\varphi^2/v_\phi^2)$, have been derived. Here, we would like to extend the flavons $\varphi$ and $\phi$ from pseudo-real scalars to complex scalars. In other words, $\varphi_1$ is not real and $\varphi_2$ is not conjugate with $\varphi_3$, and the same holds for $\phi$. The case of complex flavons is widely used in flavour model building. One example is that in various supersymmetric flavour models, flavons have to be complex to be consistent with supersymmetry. Another example is that, in some multi-Higgs models \cite{3Higgs_lepton, 3Higgs_quark}, the flavon coupling to gauge doublet $\ell_L$ is formed by three Higgs fields, which are gauge doublets and thus have to be complex. 

In the following, we assume $\varphi$ and $\phi$ to be neutral complex scalars \footnote{If $\varphi$ is a gauge doublet, the following discussion applies to its neutral components.} and transform as triplets of $A_4$, taking the same charges as in Table \ref{tab:fields}. Note that, in the basis in Eq. \eqref{eq:generators}, as $\varphi=(\varphi_1,\varphi_2,\varphi_3)$ is a triplet, $\tilde{\varphi}=(\varphi_1^*, \varphi_3^*, \varphi_2^*)$ is also a triplet. To simplify our discussion, we require $\varphi$ and $\tilde{\varphi}$ to appear in pairs of the combination $\tilde{\varphi} \varphi$, and the same applies for $\phi$ and $\tilde{\phi}$. This is easy to be achieved by imposing an additional $U(1)$ symmetry with $\varphi$ and $\phi$ taking different charges $q_\varphi$ and $q_\phi$, respectively. On the other hand, if $\varphi$ is a gauge doublet, $\varphi$ and $\tilde{\varphi}$ always appears in pairs, and no additional symmetry is needed. 

Also in this case we allow cross couplings in the potential and we study how the VEVs are modified compared to the case of no cross couplings. 
The most general potential of $\varphi$ is altered to \cite{3Higgs_lepton}
\begin{eqnarray}
V(\varphi)&=&\frac{1}{2}\mu^2_\varphi (\tilde{\varphi} \varphi)_\mathbf{1} + \frac{1}{4} \Big[ f_1 \big( (\tilde{\varphi} \varphi)_\mathbf{1}\big)^2 + f_2 (\tilde{\varphi} \varphi)_{\mathbf{1}'} (\tilde{\varphi} \varphi)_{\mathbf{1}''} + f_3 \big( (\tilde{\varphi} \varphi)_{\mathbf{3}_S} (\tilde{\varphi} \varphi)_{\mathbf{3}_S} \big)_\mathbf{1}  \nonumber\\
&& + f_4 \big( (\tilde{\varphi} \varphi)_{\mathbf{3}_A} (\tilde{\varphi} \varphi)_{\mathbf{3}_A} \big)_\mathbf{1} + f_5 \big( (\tilde{\varphi} \varphi)_{\mathbf{3}_S} (\tilde{\varphi} \varphi)_{\mathbf{3}_A} \big)_\mathbf{1} \Big] \,,
\label{eq:VPhi} 
\end{eqnarray} 
where $f_i$ are all real couplings. The additional $f_4$ and $f_5$ terms originate from the complex property of $\varphi$.  Replacing $\varphi\to\phi$ and the coefficients $f_i\to g_i$, we obtain the potential for $\phi$, $V(\phi)$. 
The cross couplings between $\varphi$ and $\phi$ are modified to
\begin{eqnarray}
\hspace{-7mm }
V(\varphi, \phi) &=& \frac{1}{2} \epsilon_1( \tilde{\varphi} \varphi )_\mathbf{1} ( \tilde{\phi} \phi )_\mathbf{1} + \frac{1}{4}\Big[ \epsilon_2 ( \tilde{\varphi} \varphi )_{\mathbf{1}''} ( \tilde{\phi} \phi )_{\mathbf{1}'} +\text{h.c.} \Big]
+ \frac{1}{2} \Big[ \epsilon_3 \big( ( \tilde{\varphi} \varphi )_{\mathbf{3}_S} (\tilde{\phi} \phi )_{\mathbf{3}_S} \big)_{\mathbf{1}} \nonumber\\
&&+ \epsilon_4 \big( ( \tilde{\varphi} \varphi )_{\mathbf{3}_A} (\tilde{\phi} \phi )_{\mathbf{3}_A} \big)_{\mathbf{1}}+ \epsilon_5 \big( ( \tilde{\varphi} \varphi )_{\mathbf{3}_S} (\tilde{\phi} \phi )_{\mathbf{3}_A} \big)_{\mathbf{1}} + \epsilon_6 \big( ( \tilde{\varphi} \varphi )_{\mathbf{3}_A} (\tilde{\phi} \phi )_{\mathbf{3}_S} \big)_{\mathbf{1}} \Big] \,,
\label{eq:Vmix2}
\end{eqnarray}
in which $\epsilon_2$ is complex and $\epsilon_1$, $\epsilon_3$, $\epsilon_4$, $\epsilon_5$ and $\epsilon_6$ are real. 

Following the procedure in the last section, we calculate the VEVs of $\varphi$ and $\phi$ perturbatively. We first consider the VEV of $\varphi$ without cross couplings. 
$\langle \varphi \rangle=(1,0,0)^Tv_{\varphi}$ and $\langle \phi \rangle=(1,1,1)^Tv_{\phi}/\sqrt{3}$ are still minima of the potential, and $V(\varphi)$ at $\langle \varphi\rangle$ takes the same value as $V(\langle \varphi\rangle_1)$ in Eq. \eqref{eq:vevs2}. 
We expand the VEVs of $\varphi$ and $\phi$ as in Eq. \eqref{eq:expansion}
and write out the quadratic terms of $V(\varphi)$, $V(\phi)$ and the linear terms of $V(\varphi,\phi)$ as 
\begin{eqnarray}
V^{(2)}(\varphi) &=& \frac{1}{2}m_{\varphi1}^2 [\text{Re}(\delta v_{\varphi1})]^2 + \frac{1}{2}
\left(\begin{array}{cc} \delta v_{\varphi2}^{*} & \delta v_{\varphi3} \end{array} \right)
O_{\varphi}\left(
\begin{array}{cc}
 m_{\varphi2}^2 & 0 \\
 0 & m_{\varphi3}^2 \\
\end{array} 
\right)O_{\varphi}^T
 \left(\begin{array}{c} \delta v_{\varphi2} \\ \delta v_{\varphi3}^{*} \end{array} \right) \,, \nonumber\\
V^{(2)}(\phi) &=& \frac{1}{2}  \left(\begin{array}{ccc} \text{Re}(\delta v_{\phi1}) & \text{Re}(\delta v_{\phi2}) & \text{Re}(\delta v_{\phi3}) \end{array}\right) 
O_{\phi}
\left(
\begin{array}{ccc}
 m_{\phi1}^2 & 0 & 0 \\
 0 & m_{\phi2}^2 & 0 \\
 0 & 0 & m_{\phi3}^2 \\
\end{array}
\right)
O_{\phi}^T 
 \left(\begin{array}{c} \text{Re}(\delta v_{\phi1}) \\ \text{Re}(\delta v_{\phi2}) \\ \text{Re}(\delta v_{\phi3}) \end{array} \right) \nonumber\\
&&+ \frac{1}{2}  \left(\begin{array}{ccc} \text{Im}(\delta v_{\phi1}) & \text{Im}(\delta v_{\phi2}) & \text{Im}(\delta v_{\phi3}) \end{array}\right) O_{\phi} 
\left(
\begin{array}{ccc}
 0 & 0 & 0 \\
 0 & m_{\phi3}^2 & 0 \\
 0 & 0 & m_{\phi2}^2 \\
\end{array}
\right)
O_{\phi}^T
 \left(\begin{array}{c} \text{Im}(\delta v_{\phi1}) \\ \text{Im}(\delta v_{\phi2}) \\ \text{Im}(\delta v_{\phi3}) \end{array} \right)\,,\nonumber\\
\hspace{-5mm}V^{(1)}(\varphi, \phi) &=&  (\epsilon_1\,\text{terms}) + \big(\frac{\epsilon_2}{4}   v_\phi^2 v_\varphi  ( \delta v_{\varphi2}^* + \delta v_{\varphi3} ) + \text{h.c.}\big) \nonumber\\
&&+ \frac{\epsilon_3}{2}  v_\varphi^2 \frac{v_\varphi}{\sqrt{3}}\big( 2\text{Re}(\delta v_{\phi1}) - \text{Re}(\delta v_{\phi2}) - \text{Re}(\delta v_{\phi3}) \big) 
+ \frac{\epsilon_5}{2}  v_\varphi^2 \frac{v_\varphi}{\sqrt{3}}\big( \text{Re}(\delta v_{\phi3}) - \text{Re}(\delta v_{\phi2}) \big)  \,, 
\label{eq:V_quadratic2}
\end{eqnarray}  
where
\begin{eqnarray} 
&&O_{\varphi}=\left(\begin{array}{cc}
 \cos\vartheta & \sin\vartheta \\
 -\sin\vartheta & \cos\vartheta \\
\end{array}\right) \,, \qquad
O_{\phi}=U_\text{TBM}
\left(\begin{array}{ccc}
 0 & \cos\theta & \sin\theta \\
 1 & 0 & 0 \\
 0 & -\sin\theta & \cos\theta \\
\end{array}\right) \,,
\nonumber\\
&&m_{\varphi1}^2=2(f_1+f_3)v_\varphi^2\,, \quad 
m_{\varphi2}^2,m_{\varphi3}^2= \left(2 f_2 - 5 f_3 + f_4 \mp \sqrt{(2 f_2 + f_3 - f_4)^2 + 4 f_5^2}\right) \frac{v_\varphi^2}{4}\,,\nonumber\\
&&m_{\phi1}^2=2(g_1+g_2)v_\phi^2\,, \quad 
m_{\phi2}^2,m_{\phi3}^2= \left(-6g_2 +3 g_3 + g_4 \mp \sqrt{(3g_3 - g_4)^2 + 3 g_5^2}\right) \frac{v_\phi^2}{4} \,,
\label{eq:flavon_mass_sub}
\end{eqnarray}
and the flavon mixing parameters $\vartheta$ and $\theta$ are respectively given by 
\begin{eqnarray}
\tan2\vartheta=\frac{2f_2+f_3-f_4}{2f_5} \,, \qquad 
\tan2\theta=\frac{\sqrt{3}g_5}{g_4-3g_3}\,,
\label{eq:theta}
\end{eqnarray} 
with $m_{\varphi2}^2\leqslant m_{\varphi3}^2$, $m_{\phi2}^2\leqslant m_{\phi3}^2$ and $-90^\circ<\vartheta,\theta\leqslant 90^\circ$ required. Here, $m_{\varphi1}^2$, $m_{\varphi2}^2$, $m_{\varphi3}^2$ and  $m_{\phi1}^2$, $m_{\phi2}^2$, $m_{\phi3}^2$ are mass eigenvalues of $\varphi$ and $\phi$ after $A_4$ is broken to $Z_3$ and $Z_2$, respectively. Note that $\varphi_2$ and $\varphi_3$ are not mass eigenstates any more since there is a mixing between $\varphi_2$ and $\varphi_3^*$, with the mixing angle characterised by $\vartheta$. 
In most cases, the mixing between $\varphi_2$ and $\varphi_3^*$ is sizable and cannot be neglected, except in the limit $2f_2+f_3-f_4 = 0$. It becomes maximal in the limit $f_5 = 0$. When both conditions $2f_2+f_3-f_4 = 0$ and $f_5 = 0$ are satisfied, the two mass eigenvalues are degenerate, being equal to $m_{\varphi2}^2$ in Model I. As for $\phi$, its real components and imaginary components mix separately after $A_4$ breaks to $Z_2$. The real components gain masses with eigenvalues $m_{\phi1}^2$, $m_{\phi2}^2$ and $m_{\phi3}^2$, while the imaginary components gain masses with eigenvalues $0$, $m_{\phi3}^2$ and $m_{\phi2}^2$. $m_{\phi2}^2$ and $m_{\phi3}^2$ can be degenerate only if both conditions $3g_3=g_4$ and $g_5=0$ hold. 
One necessary condition for the stable vacuum is that mass squares of all massive scalars are positive, which leads to \footnote{If $\varphi$ is formed by three Higgses, the squares of the charged Higgs masses are given by $m_{\varphi_2^+}^2, m_{\varphi_3^+}^2=(-6 f_3 \mp f_5) {v_\varphi^2}/{4}$. In order to avoid a charged vacuum, additional requirements $f_3<0$ and $6 f_3<f_5<-6f_3$ should be imposed. }
\begin{eqnarray}
&f_1+f_3>0\,,\quad
2f_2-5f_3+f_4>0\,,\quad
2(f_2-f_3)(f_4-3f_3)-f_5^2>0\,, \nonumber\\
&g_1+g_2>0\,,\quad
3g_3-6g_2+g_4>0\,,\quad
4(g_2-g_3)(3g_2-g_4)-g_5^2>0\,. 
\label{eq:stable_vacuum}
\end{eqnarray} 
The terms in the cross couplings which can shift the direction of the VEV at first order are only those related to $\epsilon_2$, $\epsilon_3$ and $\epsilon_5$, and in Eq. \eqref{eq:V_quadratic2} we list the linear terms for them. The $\epsilon_1$ terms can only give overall corrections to $v_\varphi$ and $v_\phi$ and can be absorbed in a redefinition of $\mu_\varphi^2$ and $\mu_\phi^2$ by performing the same procedure as described in section \ref{sec:2}. 

We minimise the potential in Eq. \eqref{eq:V_quadratic2} and directly obtain the following corrections: 
\begin{eqnarray}
\left(\begin{array}{c} \delta v_{\varphi2} \\ \delta v_{\varphi3}^* \end{array} \right) 
&\approx&
 - \frac{1}{2} v_\phi^2 v_\varphi 
O_{\varphi}\left(
\begin{array}{cc}
 m_{\varphi2}^{-2} & 0 \\
 0 & m_{\varphi3}^{-2} \\
\end{array} 
\right)O_{\varphi}^T
 \left(\begin{array}{c} \epsilon_2 \\ \epsilon_2 \end{array} \right) \,, \nonumber\\
\left(\begin{array}{c} \text{Re}(\delta v_{\varphi1}) \\ \text{Re}(\delta v_{\varphi2}) \\ \text{Re}(\delta v_{\varphi3}) \end{array} \right) 
&\approx&
 - \frac{1}{2}  v_\varphi^2 \frac{v_\varphi}{\sqrt{3}}
 O_{\phi}
\left(
\begin{array}{ccc}
 m_{\phi1}^{-2} & 0 & 0 \\
 0 & m_{\phi2}^{-2} & 0 \\
 0 & 0 & m_{\phi3}^{-2} \\
\end{array}
\right)
O_{\phi}^T 
 \left(\begin{array}{c} 2\epsilon_3 \\ -\epsilon_3 - \epsilon_5 \\ -\epsilon_3 +\epsilon_5 \end{array} \right)\,.
\end{eqnarray}
Then, the vacuum shifts for $\varphi$ and $\phi$ can be expressed as
\begin{eqnarray}
\langle \varphi \rangle \approx \begin{pmatrix}1 \\ (1-\kappa_\varphi)\epsilon_{\varphi} \\ (1+\kappa_\varphi)\epsilon_{\varphi}^* \end{pmatrix} v_\varphi  \,,\quad
\langle \phi \rangle \approx \begin{pmatrix}1-2\epsilon_\phi \\ 1+(1-\kappa_\phi)\epsilon_{\phi} \\ 1+(1+\kappa_\phi)\epsilon_{\phi} \end{pmatrix} \frac{v_\phi}{\sqrt{3}} \,,
\end{eqnarray}
respectively with 
\begin{eqnarray} 
\epsilon_{\varphi} &=& -\frac{ \epsilon_2 v_\phi^2 \big[(m_{\varphi3}^2 + m_{\varphi2}^2) - (m_{\varphi3}^2 - m_{\varphi2}^2) \sin2\vartheta \big]}{4 m_{\varphi3}^2 m_{\varphi2}^2} \,, \nonumber\\
\kappa_\varphi &=& \frac{ (m_{\varphi3}^2 - m_{\varphi2}^2) \cos2\vartheta }{ (m_{\varphi3}^2 + m_{\varphi2}^2) - (m_{\varphi3}^2 - m_{\varphi2}^2) \sin2\vartheta} \,,\nonumber\\
\epsilon_{\phi} &=& \frac{ v_\varphi^2 \big[\epsilon_3(m_{\phi3}^2 + m_{\phi2}^2) + (m_{\phi3}^2 - m_{\phi2}^2) (\epsilon_3 \cos2\theta - \frac{\epsilon_5}{\sqrt{3}}\sin 2\theta) \big]}{4 m_{\phi3}^2 m_{\phi2}^2} \,, \nonumber\\
\kappa_\phi &=& \frac{ (m_{\phi3}^2 - m_{\phi2}^2) (\sqrt{3} \epsilon_3 \sin2\theta - \epsilon_5 \cos 2\theta) - \epsilon_5 (m_{\phi3}^2 + m_{\phi2}^2)}{ \epsilon_3 (m_{\phi3}^2 + m_{\phi2}^2) + (m_{\phi3}^2 - m_{\phi2}^2) (\epsilon_3 \cos2\theta - \frac{\epsilon_5}{\sqrt{3}}\sin 2\theta)} \,.
\label{eq:epsilon_kappa}
\end{eqnarray} 
These expressions reflect the complex properties of $\varphi$ and $\phi$. 
When the mixing between $\varphi_2$ and $\varphi_3^{*}$ is maximal ($\sin2\vartheta=\pm1$, corresponding to $f_5=0$), $\kappa_\varphi$ vanishes, and we get the $\langle \varphi \rangle$ shift similar to that in Model I. Furthermore, $\epsilon_\varphi$ takes the value $- \epsilon_2 v_\phi^2/(2m_{\varphi3}^2)$  and $- \epsilon_2 v_\phi^2/(2m_{\varphi2}^2)$
for $\vartheta=45^\circ$ and $-45^\circ$, respectively. 
In the limit $\sin2\theta=0$ (corresponding to $g_5=0$), $\kappa_\phi$ vanishes, and we also get the $\langle \phi \rangle$ shift similar to that in Model I, with $\epsilon_\phi=\epsilon_3 v_\varphi^2/(2 m_{\phi2}^2)$ for $\theta=0$ and $\epsilon_\phi=\epsilon_3 v_\varphi^2/(2 m_{\phi3}^2)$ for $\theta=90^\circ$. 

Let us consider the corrections inducing to the TBM pattern. In $A_4\times Z_2 \times Z_4$, all fields are assumed to take the same charges as in Model I. With a suitable arrangement of the $U(1)$ charge for these fields, couplings such as $(\overline{\ell_L} \tilde{\varphi})_\mathbf{1} e_R H$ and $\big((\overline{N^c} N)_{\mathbf{3}_S} \tilde{\phi} \big)_\mathbf{1}$ can be forbidden. Eventually, we can obtain the same Lagrangian in Eq. \eqref{eq:Lagrangian} \footnote{If $\varphi$ is a gauge doublet, we can construct a renormalisable model with terms generating charged lepton masses such as 
$$
-\mathcal{L}_l = y_e (\overline{\ell_L} \varphi)_\mathbf{1} e_R + y_\mu (\overline{\ell_L} \varphi)_{\mathbf{1}''} \mu_R + y_\tau (\overline{\ell_L} \varphi)_{\mathbf{1}'} \tau_R + \text{h.c.}  \,. $$
}. 
After the scalars get VEVs,
the charged lepton and right-handed neutrino mass matrices are given by
\begin{eqnarray}
M_l&\approx&\left(
\begin{array}{ccc}
 y_e &  y_\mu (1+\kappa_\varphi) \epsilon_{\varphi}^* &  y_\tau (1-\kappa_\varphi)\epsilon_{\varphi} \\
 y_e(1-\kappa_\varphi) \epsilon_{\varphi} &  y_\mu &  y_\tau (1+\kappa_\varphi) \epsilon_{\varphi}^* \\
 y_e (1+\kappa_\varphi) \epsilon_{\varphi}^* &  y_\mu (1-\kappa_\varphi) \epsilon_{\varphi} &  y_\tau \\
\end{array}
\right)\frac{ v v_\varphi}{\sqrt{2}\Lambda} \,,\nonumber\\
M_N&\approx& \frac{1}{\sqrt{3}}
\left(
\begin{array}{ccc} 
\sqrt{3}  y_2v_\eta+ y_1v_\phi [1-2 \epsilon_\phi] & - \frac{1}{2} y_1v_\phi [1+(1+\kappa_\phi)\epsilon_\phi] & - \frac{1}{2} y_1v_\phi [1+(1-\kappa_\phi)\epsilon_\phi] \\
 - \frac{1}{2} y_1v_\phi [1+(1+\kappa_\phi)\epsilon_\phi] & y_1v_\phi [1+(1-\kappa_\phi)\epsilon_\phi] & \sqrt{3} y_2v_\eta- \frac{1}{2} y_1v_\phi [1-2 \epsilon_\phi] \\
 - \frac{1}{2} y_1v_\phi [1+(1-\kappa_\phi)\epsilon_\phi] & \sqrt{3} y_2v_\eta- \frac{1}{2} y_1v_\phi [1-2 \epsilon_\phi] & y_1v_\phi [1+(1+\kappa_\phi)\epsilon_\phi] \\
\end{array}
\right)\,,
\label{eq:lepton_mass2}
\end{eqnarray}
and the Dirac neutrino mass matrix takes the same form as in Eq. \eqref{eq:neutrino_mass}. The lepton mass eigenvalues are the same as those in Model I. 
The PMNS matrix is given by
\begin{eqnarray}
\hspace{-5mm} 
U_\text{PMNS}\approx\left(
\begin{array}{ccc}
 1 & -(1+\kappa_\varphi)\epsilon_{\varphi}^* & -(1-\kappa_\varphi)\epsilon_{\varphi} \\
(1+\kappa_\varphi) \epsilon_{\varphi} & 1 & -(1+\kappa_\varphi)\epsilon_{\varphi}^* \\
 (1-\kappa_\varphi)\epsilon_{\varphi}^* & (1+\kappa_\varphi)\epsilon_{\varphi} & 1 \\
\end{array}
\right) \!\!
\left(
\begin{array}{ccc}
 \frac{2}{\sqrt{6}} & \frac{1}{\sqrt{3}} & 0 \\
 -\frac{1}{\sqrt{6}} & \frac{1}{\sqrt{3}} & -\frac{1}{\sqrt{2}} \\
 -\frac{1}{\sqrt{6}} & \frac{1}{\sqrt{3}} & \frac{1}{\sqrt{2}} \\
\end{array}
\right)\!\!
\left(
\begin{array}{ccc}
 1 & \!\!\!\sqrt{2}\epsilon_\phi\!\!\! & 0 \\
 \!\!-\sqrt{2} \epsilon_\phi\!\!\! & 1 & \!\!\!\sqrt{6} \kappa_\phi' \epsilon_\phi\!\! \\
 0 & \!\!\!-\sqrt{6} \kappa_\phi' \epsilon_\phi\!\!\! & 1 \\
\end{array}
\right)\! P_\nu \,, 
\label{eq:PMNS2}
\end{eqnarray}
where 
\begin{eqnarray}
\kappa_\phi'=\frac{\kappa_\phi y_1 v_\phi}{3y_1 v_\phi - 4\sqrt{3} y_2 v_\eta } \,.
\end{eqnarray}
Eventually, we obtain approximate expressions for the mixing angles 
\begin{eqnarray}
&&\sin\theta_{13}\approx \sqrt{ 2 |\epsilon_\varphi|^2\sin^2 \theta_\varphi + 2 ( \kappa_\varphi|\epsilon_\varphi|\cos\theta_\varphi + \kappa_\phi^{\prime}\epsilon_\phi )^2 }\,,\nonumber\\
&&\sin\theta_{12}\approx \frac{1}{\sqrt{3}}\big[1-2|\epsilon_\varphi|\cos\theta_\varphi+2\epsilon_\phi\big]\,,\nonumber\\
&&\sin\theta_{23}\approx \frac{1}{\sqrt{2}}\big[1+(1+\kappa_\varphi)|\epsilon_\varphi| \cos\theta_\varphi -2\kappa_\phi' \epsilon_\phi \big] \,,
\label{eq:mixing_angles2}
\end{eqnarray}
and the CP-violating phases
\begin{eqnarray}
&&\delta\approx\text{Arg} \Big\{ \Big[ i |\epsilon_\varphi|\sin\theta_\varphi + \kappa_\varphi |\epsilon_\varphi|\cos\theta_\varphi + \kappa_\phi' \epsilon_\phi \Big] \Big[ -1+ i (2+\kappa_\varphi) |\epsilon_\varphi| \sin\theta_\varphi  \Big] \Big\} \,, \nonumber\\
&& \alpha_{21}\approx\text{Arg}\left\{ \Big[1+\frac{\sqrt{3} y_1 v_\phi}{ 2 y_2 v_\eta}\Big]\Big[1-6i \kappa_\varphi|\epsilon_\varphi|\sin\theta_\varphi\Big] \right\} \,, \nonumber\\
&& \alpha_{31}\approx\text{Arg}\left\{ \Big[ \frac{\sqrt{3} y_1 v_\phi + 2 y_2 v_\eta}{\sqrt{3} y_1 v_\phi - 2 y_2 v_\eta} \Big] \Big[ 1- 4 i |\epsilon_\varphi| \sin\theta_\varphi \Big] \right\}\,.
\label{eq:CP_phases2} 
\end{eqnarray} 
Compared with Model I, two additional parameters $\kappa_\varphi$ and $\kappa_\phi$, which stand for the asymmetric corrections between the second and third components of the flavon VEVs, come into the game. They result in some different features compared to Model I. Here we discuss two such possibilities: 

\begin{figure}[t!]
\begin{center}
\includegraphics[width=1\textwidth,natwidth=600,natheight=600]{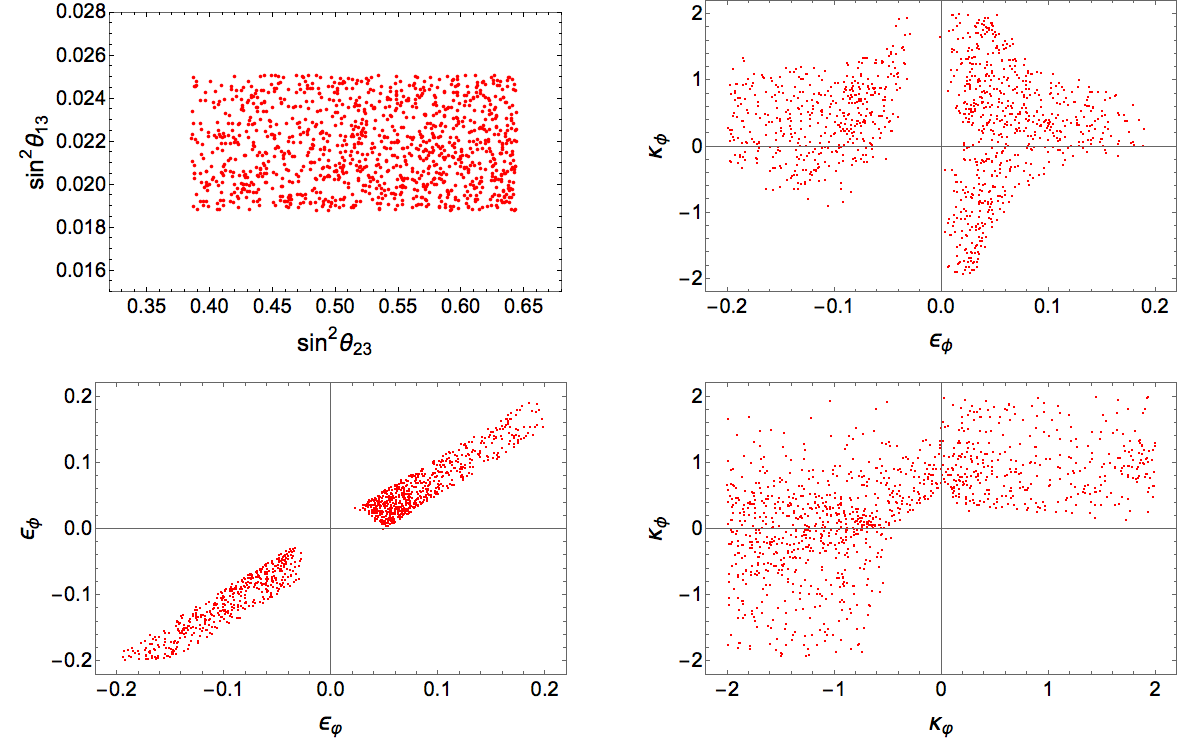}
\vspace{-1cm} 
\caption{Theoretical prediction of mixing angles (left top panel) and the allowed parameter space of the parameters $\epsilon_\varphi$, $\epsilon_{\phi}$ and $\kappa_\varphi$, $\kappa_\phi$ (top right and bottom panels) in Case A, Model II. The $3\sigma$ ranges of mixing angles in Ref. \cite{globalfit} are used as inputs. $\epsilon_\varphi$ is assumed to be real. $\epsilon_\varphi$, $\epsilon_\phi$ and $\kappa_\varphi$, $\kappa_{\phi}$ are samples randomly generated in the bounds $[-0.2,\,0.2]$ and $[-2,\,2]$, respectively, and the points compatible with data are shown in the figure.} 
\label{fig:mixII}
\end{center}
\end{figure}

\begin{figure}[t!]
\begin{center}
\includegraphics[width=1\textwidth,natwidth=600,natheight=600]{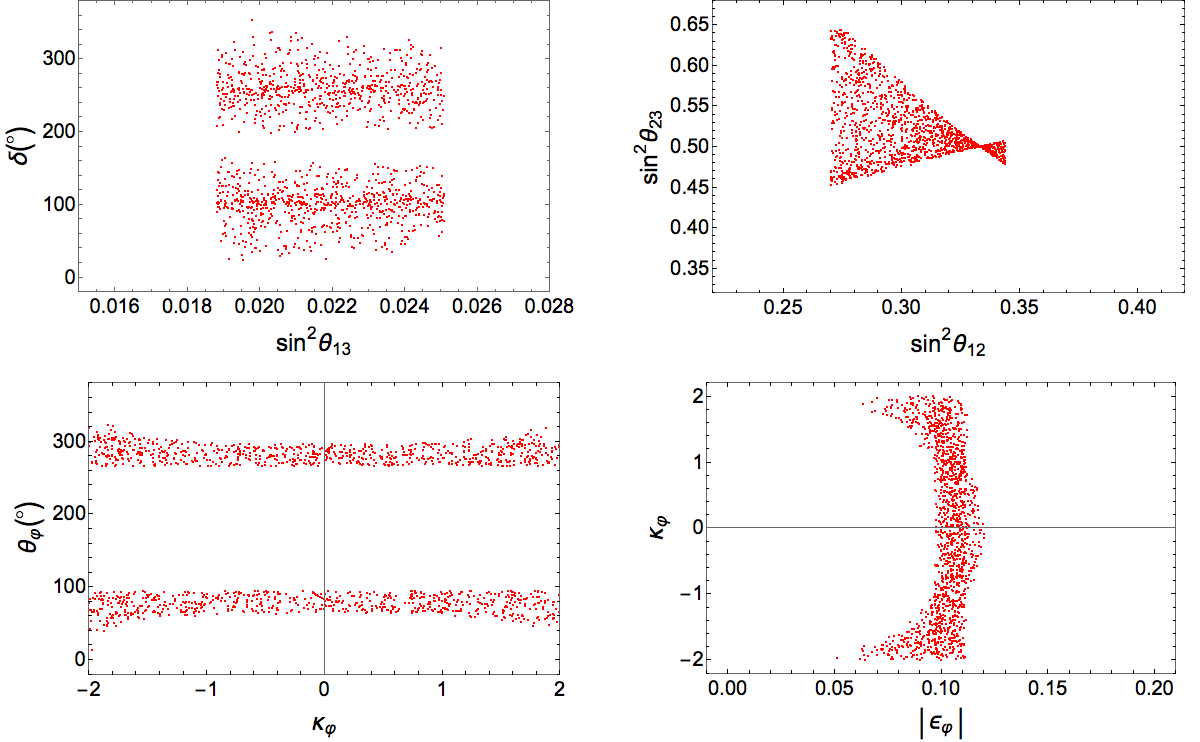}
\vspace{-1cm} 
\caption{Theoretical predictions of mixing angles (top panels) and parameter spaces of the parameters $\epsilon_\varphi$, $\epsilon_{\phi}$ and $\kappa_\varphi$, $\kappa_\phi$ (bottom panels) in Case B, Model II, where $\epsilon_\phi=0$, and $|\epsilon_\varphi|$, $\kappa_\varphi$ and $\theta_{\varphi}$ are samples randomly generated in the ranges $[0,\,0.2]$, $[-2,\,2]$ and $[0^\circ,\,360^\circ)$, respectively, and the points compatible with data are shown in the figure. The same inputs of mixing angles have been employed as in Fig. \ref{fig:mixII}. }
\label{fig:mixIII}
\end{center}
\end{figure}
\begin{itemize} 

\item Case A: All coefficients in the potential $V(\varphi, \phi)$ are real, i.e., $\epsilon_\varphi$ being real. In this case, there will be no Dirac-type CP violation, and the corrections to the mixing angles are simplified to
\begin{eqnarray}
r'=|\kappa_\varphi \epsilon_\varphi + \kappa_\phi'\epsilon_\phi |\,,\quad
s=2\epsilon_\phi-2\epsilon_\varphi \,,\quad
a=(1+\kappa_\varphi) \epsilon_\varphi - 2 \kappa_\phi' \epsilon_\phi \,. 
\label{eq:caseA_rsa}
\end{eqnarray}
As seen from the expression of $r^\prime$, the reactor mixing angle depends on two additional parameters, the asymmetric shifts of the VEVs $\langle \varphi \rangle$ and $\langle \phi \rangle$, characterised by $\kappa_\varphi$ and $\kappa_\phi'$, respectively. Each can be related to the asymmetric coupling $f_5 \big( (\tilde{\varphi} \varphi)_{\mathbf{3}_S} (\tilde{\varphi} \varphi)_{\mathbf{3}_A} \big)_\mathbf{1}$ or $g_5 \big( (\tilde{\phi} \phi)_{\mathbf{3}_S} (\tilde{\phi} \phi)_{\mathbf{3}_A} \big)_\mathbf{1}$, see Eqs. \eqref{eq:theta} and \eqref{eq:epsilon_kappa}. These couplings also contribute to the correction to $\theta_{23}$, but not to $\theta_{12}$. In Fig. \ref{fig:mixII}, we show the prediction of $\theta_{13}$ vs $\theta_{23}$ and the allowed parameter spaces of $\epsilon_\varphi$ vs $\epsilon_{\phi}$ and $\kappa_\varphi$ vs $\kappa_\phi$. We take the $3\sigma$ allowed ranges of mixing angles in Ref. \cite{globalfit} as inputs and treat $\epsilon_{\varphi}$ and $\epsilon_{\phi}$ as random numbers in the range $[-0.2,\,0.2]$ and $\kappa_\varphi$ and $\kappa_\phi$ in the range $[-2,\,2]$. $\kappa_\varphi$ and $\kappa_\phi$ can take any values from $-\infty$ to $\infty$ in principle.  Numerically, we have checked that given random values of $m_{\varphi2}$, $m_{\varphi3}$ and $\vartheta$ and assuming that there is no large hierarchy between $\epsilon_3$ and $\epsilon_5$ ($|\epsilon_5/\epsilon_3|\in[1/2,2]$), most values of $\kappa_\varphi$ and $\kappa_\phi$ are located in the range $[-2,\,2]$. There is no strong correlation between mixing angles, and the predictions for all three mixing angles are almost evenly distributed in their $3\sigma$ ranges, except for a very weak preference for the second octant of $\theta_{23}$ as shown in Fig.~\ref{fig:mixII}. Sizable $\epsilon_\phi,\epsilon_\varphi\gtrsim0.1$ with $\epsilon_\phi\approx \epsilon_\varphi$ are allowed by constraints. They can give rise to a sizable $\theta_{13}$ and avoid a large correction to $\theta_{12}$. If $\epsilon_\varphi$ and $\epsilon_\phi$ are small, a relatively large $\kappa_\varphi$ or $\kappa_\phi$ will be preferred to give a sufficiently large correction to $\theta_{13}$. 
Finally, we note that in this model, one cannot assume that all corrections come from the neutrino sector. Otherwise we will arrive at $|a|=2r'=2|\kappa_\phi'\epsilon_\phi |$ from Eq. \eqref{eq:caseA_rsa}, and this is not compatible with current constraints on $a$ and $r'$, e.g., $|a| \lesssim 0.1$ and $r'\approx 0.1$. 

\item Case B: The correction in the neutrino sector is much smaller than the correction in the charged lepton sector. As discussed in Model I, this happens when $v_\phi$ is significantly larger than $v_\varphi$. After $\epsilon_\phi$ is neglected, 
the corrections to the mixing angles are simplified to
\begin{eqnarray}
r'=|\epsilon_\varphi| \sqrt{ \sin^2\theta_\varphi + \kappa_\varphi^2 \cos^2\theta_\varphi }\,,\quad
s=2 |\epsilon_\varphi| \cos\theta_\varphi \,,\quad
a=(1+\kappa_\varphi) |\epsilon_\varphi| \cos\theta_\varphi \,. 
\end{eqnarray}
The correlations of mixing parameters and the allowed parameter space of $|\epsilon_\varphi|$, $\theta_{\varphi}$ and $\kappa_\varphi$ are shown in Fig. \ref{fig:mixIII}. The Dirac phase $\delta$ may deviate from $90^\circ$ or $270^\circ$ greatly and take a value in the range $[0,\,360^\circ)$. However, in most cases, it takes a value in the range $(50^\circ,150^\circ)$ or $(210^\circ,310^\circ)$. The sum rule $|\epsilon_\varphi|^2=r^{\prime2}+a(s-a)$ is satisfied. To be compatible with data, the value of $|\epsilon_\varphi|$ is in general around 0.1. A small $\cos\theta_\varphi$ is also preferred, similar to that in Model I, which allows $|\epsilon_\varphi|\sin\theta_\varphi$ to give a sizable correction to $\theta_{13}$ and not significantly modify $\theta_{12}$ and $\theta_{23}$. In detail, $\theta_\varphi$ is mostly constrained in the ranges $(50^\circ,\,100^\circ)$ and $(270^\circ,\,310^\circ)$ in this model. 

\end{itemize}

\section{Conclusion and discussion \label{sec:4}}

Flavons play a key role in leptonic flavour models with discrete flavour symmetries. They gain vacuum expectation values (VEVs), breaking the high energy flavour symmetry and leaving residual symmetries different in the charged lepton ($\varphi$) and in the neutrino ($\phi$) sectors. This misalignment leads to specific  flavour mixing structures. 
In most models, $\theta_{13}$ vanishes and the CP invariance is conserved at leading order. In order to be compatible with observations, higher dimensional operators are typically introduced to modify $\theta_{13}$, $\theta_{12}$, $\theta_{23}$ from their leading order values and induce $\delta \neq 0,\pi$. 
In this paper, we exploit a different approach in which we emphasise the importance of flavon cross couplings to flavour mixing. We find that cross couplings between different flavons can break the residual symmetries, shifting the VEVs of flavons and modifying flavour mixing. These couplings provide new origins for the non-zero $\theta_{13}$ and CP violation. 

For definiteness, we present two models based on $A_4$. Depending on the coefficients in the flavon potential, different vacua preserving different residual symmetries can be identified. By appropriately choosing them in the charged lepton and neutrino sectors, we can realise the tri-bimaximal (TBM) mixing at leading order. The cross couplings between different flavons result in the breaking of the residual symmetries and corrections to TBM. 

In Model I, both flavons $\varphi$ and $\phi$ are assumed to be pseudo-real triplets of $A_4$. The cross coupling $\text{Im}(\epsilon_2) ( \varphi \varphi )_{\mathbf{1}''} ( \phi \phi )_{\mathbf{1}'}$ leads to the vacuum shift of $\varphi$ and the breaking of the $Z_3$ residual symmetry in the charged lepton sector, where the relative size of the breaking is characterised by a complex parameter $\epsilon_\varphi$. Both $\delta$ and $\theta_{13}$ arise from this term and consequently are connected by a sum rule $\delta=270^\circ-\sqrt{2}\theta_{13}$. Taking account of current oscillation data, we predict $\delta\approx 258^\circ$, very close to the current best-fit value. The flavon cross couplings also lead to the breaking of the $Z_2$ residual symmetry in the neutrino sector, characterised by a real parameter $\epsilon_\phi$. The solar angle $\theta_{12}$ and the atmospheric angle $\theta_{23}$ gain corrections from both $\epsilon_\varphi$ and $\epsilon_\phi$. In the interesting case in which the VEV $v_\phi$ is significantly greater than $v_\varphi$, the correction $\epsilon_\phi$ is negligible compared with $\epsilon_\varphi$ since $\epsilon_\phi/\epsilon_\varphi$ is suppressed by $v_\varphi^4 / v_\phi^4$. All modifications to TBM arise from one single parameter $\epsilon_\varphi$, and an additional sum rule $3\sin^2\theta_{12} + 4\sin^2\theta_{23} =3$ is obtained. 

In Model II, flavons are assumed to be complex scalars. This extension is natural due to the consistency with supersymmetric models and multi-Higgs models and brings some new features which are absent in Model I. One is that it provides new sources for non-zero $\theta_{13}$. Due to the complex property of the flavons, some asymmetric couplings are included in the flavon potential and they lead to asymmetric modifications between the second and third components of the flavon VEVs, parametrised by $\kappa_\varphi$ and $\kappa_\phi$. The latter can induce sizable $\theta_{13}$ while not affecting CP conservation in some specific region of the parameter space. If the correction in the neutrino sector is negligibly small compared with that in the charged lepton sector, the Dirac phase prefers to take a value not far from maximal CP violation. 

The flavon couplings and in particular the cross couplings can have other phenomenological consequences, which, depending on the flavon mass scales, can be tested directly. They may be at the origin of other types of lepton flavour violation, namely lepton-flavour-violating decays of charged leptons such as $\mu\to e\gamma$ and $\tau \to ee e $. Another type of cross couplings which we have not considered here are those between flavons and Higgs. Although not relevant for the leptonic flavour structure, such couplings provide ways to detect flavons directly and indirectly at colliders, through, e.g., the associated production with the Higgs and precision measurement of the couplings of the Higgs, respectively. Detailed studies of these aspects will be carried out in the future. 

In conclusion, we have shown that cross couplings between different flavons may be the origin of the reactor mixing angle and CP violation. This is a new way, different from higher dimensional operators, to modify flavour mixing from its leading order result. Very few degrees of freedom are introduced in models based on cross couplings, which makes them much simpler than those built in the framework of extra dimension or supersymmetry. The approach proposed in this paper is not limited in $A_4$ models, and can be easily applied into other models with different discrete flavour symmetries. 

\section*{Acknowledgement}

We would like to thank Dr. Peter Ballett for his useful advice. YLZ is also grateful to Prof. Zhi-zhong Xing for useful discussions and to Prof. Wei Wang for warm hospitality at Sun Yat-Sen University. 
This work was supported by the European Research Council under ERC Grant “NuMass” (FP7-IDEAS-ERC ERC-CG 617143), and by the European Union FP7 ITN-INVISIBLES (Marie Curie Actions, PITN-GA-2011-289442).

\appendix
\section{The vacuum alignment for a pseudo-real triplet flavon of $A_4$ \label{sec:A}}

We calculate the vacuum of the $A_4$-invariant potential $V(\varphi)$ in Eq. \eqref{eq:Vvarphi} in the Ma-Rajasekaran basis \cite{Ma:2001dn}. This basis is easier for us to find out all vacuums of $\varphi$  than the Altarelli-Feruglio basis \cite{Altarelli:2005yx}, although physics is equivalent in different bases. After we find out the solutions, we will rotate them to the Altarelli-Feruglio basis, in which charged lepton mass matrix is nearly diagonal. 

In the Ma-Rajasekaran basis, generators of $A_4$ in the 3d irreducible represention are written as
\begin{eqnarray}
T=\left(
\begin{array}{ccc}
 0 & 0 & 1 \\
 1 & 0 & 0 \\
 0 & 1 & 0 \\
\end{array}
\right)\,,\quad
S=\left(
\begin{array}{ccc}
 1 & 0 & 0 \\
 0 & -1 & 0 \\
 0 & 0 & -1 \\
\end{array}
\right)\,.
\label{eq:generator1}
\end{eqnarray} 
The Kronecker product for two triplets $a=(a_1, a_2,a_3)^T$ and $b=(b_1, b_2, b_3)^T$ is divided into the following irreducible representations: 
\begin{eqnarray}
(ab)_\mathbf{1}\;\, &=& a_1b_1 + a_2b_2 + a_3b_3 \,,\nonumber\\
(ab)_\mathbf{1'}\, &=& a_1b_1 + \omega a_2b_2 + \omega^2 a_3b_3 \,,\nonumber\\
(ab)_\mathbf{1''} &=& a_1b_1 + \omega^2 a_2b_2 + \omega a_3b_3 \,,\nonumber\\
(ab)_{\mathbf{3}_S} &=& \frac{\sqrt{3}}{2}(a_2b_3+a_3b_2, a_3b_1+a_1b_3, a_1b_2+a_2b_1)^T \,,\nonumber\\
(ab)_{\mathbf{3}_A} &=& \frac{i}{2} (a_2b_3-a_3b_2, a_3b_1-a_1b_3, a_1b_2-a_2b_1)^T \,.
\end{eqnarray} 
In this basis, the flavon triplet $\varphi$ which is pseudo-real in the Altarelli-Feruglio basis becomes real, $\varphi_i^*=\varphi_i$ (for $i=1,2,3$) and the flavon potential $V(\varphi)$ take a simple form
\begin{eqnarray}
V(\varphi)=\frac{1}{2}\mu_\varphi^2 (\varphi_1^2+\varphi_2^2+\varphi_3^2) + \frac{1}{4}(f_1+f_2) (\varphi_1^2+\varphi_2^2+\varphi_3^2)^2+\frac{3}{4}(f_3-f_2) (\varphi_1^2\varphi_2^2+\varphi_2^2\varphi_3^2+\varphi_3^2\varphi_1^2).
\end{eqnarray}
In order to achieve a nontrivial and stable vacuum, we require a negative-definite quadratic term and a positive-definite quartic term in $V(\varphi)$, and this leads to $\mu_\varphi^2<0$, and $f_1+f_2, f_1+f_3>0$,  respectively. 
A necessary condition for the vacuum of $\varphi$ is $\partial V(\varphi)/\partial \varphi_i=0$, which is expressed as
\begin{eqnarray}
\frac{\partial V(\varphi)}{\partial \varphi_i}=\varphi_i\left[\mu_\varphi^2+(f_1+f_2)\varphi_i^2+\frac{1}{2}(2f_1-f_2+3f_3)(\varphi_j^2+\varphi_k^2)\right]=0
\end{eqnarray} 
for $i,j,k=1,2,3$ and $i\neq j \neq k \neq i$. One can obtain all solutions from the above equations directly.  These solutions are divided into three classes, according to the corresponding values of $V(\varphi)$: 
\begin{eqnarray}
\text{(1)}&&\langle \varphi \rangle_1=\left\{\begin{pmatrix} 1 \\ 1 \\ 1 \end{pmatrix},~ \begin{pmatrix} -1 \\ 1 \\ 1 \end{pmatrix},~\begin{pmatrix} 1 \\ -1 \\ 1 \end{pmatrix},~\begin{pmatrix} 1 \\ 1 \\ -1 \end{pmatrix}\right\}\frac{v_{\varphi1}}{\sqrt{3}}\,,\quad\nonumber\\
&&\hspace{4cm} v_{\varphi1}^2=\frac{-\mu_\varphi^2}{f_1+f_3}\,, \hspace{2.2cm}
V(\langle \varphi \rangle_1)=-\frac{\mu_\varphi^4}{4(f_1+f_3)}\,,\nonumber\\
\text{(2)}&&\langle \varphi \rangle_2=\left\{\begin{pmatrix} 1 \\ 0 \\ 0 \end{pmatrix}, ~\begin{pmatrix} 0 \\ 1 \\ 0 \end{pmatrix},~\begin{pmatrix} 0 \\ 0 \\ 1 \end{pmatrix}\right\}v_{\varphi2}\,,\quad\nonumber\\
&&\hspace{4cm} v_{\varphi2}^2=\frac{-\mu_\varphi^2}{f_1+f_2}\,,  \hspace{2.2cm}
V(\langle \varphi \rangle_2)=-\frac{\mu_\varphi^4}{4(f_1+f_2)}\,,\nonumber\\
\text{(3)}&&\langle \varphi \rangle_3=\left\{\begin{pmatrix} 0 \\ 1 \\ 1 \end{pmatrix},~ \begin{pmatrix} 1 \\ 0 \\ 1 \end{pmatrix},~ \begin{pmatrix} 1 \\ 1 \\ 0 \end{pmatrix},~ \begin{pmatrix} 0 \\ 1 \\ -1 \end{pmatrix},~\begin{pmatrix} -1 \\ 0 \\ 1 \end{pmatrix},~ \begin{pmatrix} 1 \\ -1 \\ 0 \end{pmatrix}\right\}\frac{v_{\varphi3}}{\sqrt{2}}\,,\quad\nonumber\\
&&\hspace{3.5cm} v_{\varphi3}^2=\frac{-4\mu_\varphi^2}{(f_1+f_2)+3(f_1+f_3)}\,,  \hspace{.3cm}
V(\langle \varphi \rangle_3)=-\frac{\mu_\varphi^4}{(f_1+f_2)+3(f_1+f_3)}\,.
\label{eq:solution1}
\end{eqnarray}
Here, each solution in the first class of solutions preserves a different $Z_3$ symmetry, e.g., $(1,1,1)^T$ invariant in a $Z_3$ generated by $T$ and $(-1,1,1)^T$ invariant in another $Z_3$ generated by $STS$. Similarly, each solution in the second class of solutions preserves a different $Z_2$ symmetry, e.g., $(1,0,0)^T$ invariant in a $Z_2$ generated by $S$ and $(0,1,0)^T$ invariant in another $Z_2$ generated by $TST^2$. 

In order to get a vacuum at $\langle \varphi \rangle_a$ (for $a=1,2,3$), we require $V(\varphi)$ take a local minimum at $\langle \varphi \rangle_a$. And this corresponds to the requirement of the positive-definite  second derivative of $V(\varphi)$. In detail, the matrix $M_\varphi^2$ defined in the following should be positive-definite at $\langle \varphi \rangle_a$: 
\begin{eqnarray}
(M_\varphi^2)_{ij}=\frac{\partial^2 V(\varphi)}{\partial \varphi_i \partial \varphi_j} \,.
\end{eqnarray}
In general, $M_\varphi^2$ is a $3\times 3$ real symmetric matrix which can be diagonalised through $W^T M_\varphi^2 W=\text{diag}\{m_{\varphi1}^2,m_{\varphi2}^2,m_{\varphi3}^2\}$, with $m_{\varphi i}^2$ the eigenvalues of $M_\varphi^2$. In the above three classes of solutions, we get
\begin{eqnarray}
\text{(1)}&&m_{\varphi1}^2=2 (f_1+f_3) v_{\varphi1}^2\,,\quad
m_{\varphi2}^2=m_{\varphi3}^2=2 (f_2-f_3) v_{\varphi1}^2\nonumber\\
\text{(2)}&&m_{\varphi1}^2=2 (f_1+f_2) v_{\varphi2}^2\,,\quad
m_{\varphi2}^2=m_{\varphi3}^2=\frac{3}{2} (f_3-f_2) v_{\varphi2}^2\nonumber\\
\text{(3)}&&m_{\varphi1}^2=\frac{1}{2} [(f_1+f_2)+3(f_1+f_3)] v_{\varphi3}^2\,,\quad
m_{\varphi2}^2=-2m_{\varphi3}^2=\frac{3}{2} (f_2-f_3) v_{\varphi3}^2\,.
\label{eq:flavon_masses}
\end{eqnarray}
at $\langle \varphi \rangle_1$, $\langle \varphi \rangle_2$ and $\langle \varphi \rangle_3$, respectively. 
Although $m_{\varphi1}^2$ is always positive in all solutions, $m_{\varphi2}^2$ or $m_{\varphi3}^2$ may be positive or negative, depending on the sign of the coefficient $f_2-f_3$. 
The third class of solutions are less interesting since $m_{\varphi2}^2$ and $m_{\varphi3}^2$ always take opposite signs at $\langle \varphi \rangle_3$. Thus, $\langle \varphi \rangle_3$ is always an unstable saddle point of $V(\varphi)$ and cannot be a vacuum. 
For the first two classes of solutions, if $f_2-f_3>0$ is required, $m_{\varphi2}^2$ and $m_{\varphi3}^2$ are positive at $\langle \varphi \rangle_1$ and negative at $\langle \varphi \rangle_2$. $\langle \varphi \rangle_2$ is an unstable saddle point. $V(\varphi)$ can only take a local minimum value (thus, also the global minimum value) at $\langle \varphi \rangle_1$. Therefore, $\langle \varphi \rangle_1$ is the only choice of the $\varphi$ VEV. On the contrary, if $f_2-f_3<0$, $\langle \varphi \rangle_2$ is the only choice of the $\varphi$ VEV.

Now we turn to the Altarelli-Feruglio basis in which the 3d irreducible generators are given in Eq. \eqref{eq:generators}.
They are obtained through a basis transformation of Eq.\eqref{eq:generator1}: $ U_\omega T U_\omega^\dag \to T$, $U_\omega S U_\omega^\dag \to S$, where 
\begin{eqnarray}
U_\omega = \frac{1}{\sqrt{3}}\left(
\begin{array}{ccc}
 1 & 1 & 1 \\
 1 & \omega ^2 & \omega  \\
 1 & \omega  & \omega ^2 \\
\end{array}
\right)\,.
\end{eqnarray}
This basis is widely implied in the literature since the charged lepton mass matrix invariant under $T$ is diagonal in this basis. A real 3d irreducible representation in the Ma-Rajasekaran basis becomes pseudo-real in this basis: $\varphi_1^*=\varphi_1$, $\varphi_2^*=\varphi_3$. The products of two 3d irreducible representations $a$ and $b$ can be expressed as
\begin{eqnarray}
(ab)_\mathbf{1}\;\, &=& a_1b_1 + a_2b_3 + a_3b_2 \,,\nonumber\\
(ab)_\mathbf{1'}\, &=& a_3b_3 + a_1b_2 + a_2b_1 \,,\nonumber\\
(ab)_\mathbf{1''} &=& a_2b_2 + a_1b_3 + a_3b_1 \,,\nonumber\\
(ab)_{\mathbf{3}_S} &=& \frac{1}{2} (2a_1b_1-a_2b_3-a_3b_2, 2a_3b_3-a_1b_2-a_2b_1, 2a_2b_2-a_3b_1-a_1b_3)^T \,,\nonumber\\
(ab)_{\mathbf{3}_A} &=& \frac{1}{2} (a_2b_3-a_3b_2, a_1b_2-a_2b_1, a_3b_1-a_1b_3)^T \,.
\label{eq:CG2}
\end{eqnarray}
Solutions in Eq. \eqref{eq:solution1} transform to
\begin{eqnarray}
\text{(1)}&&\langle \varphi \rangle_1=\left\{\begin{pmatrix} 1 \\ 0 \\ 0 \end{pmatrix}, \begin{pmatrix} \frac{1}{3} \\ -\frac{2}{3} \\ -\frac{2}{3} \end{pmatrix},\begin{pmatrix} \frac{1}{3} \\ -\frac{2}{3}\omega^2 \\ -\frac{2}{3}\omega \end{pmatrix},\begin{pmatrix} \frac{1}{3} \\ -\frac{2}{3}\omega \\ -\frac{2}{3}\omega^2 \end{pmatrix}\right\}v_{\varphi1}\,,\quad\nonumber\\
\text{(2)}&&\langle \varphi \rangle_2=\left\{\begin{pmatrix} 1 \\ 1 \\ 1 \end{pmatrix}, \begin{pmatrix} 1 \\ \omega^2 \\ \omega \end{pmatrix},\begin{pmatrix} 1 \\ \omega \\ \omega^2 \end{pmatrix}\right\}\frac{v_{\varphi2}}{\sqrt{3}}\,,\quad\nonumber\\
\text{(3)}&&\langle \varphi \rangle_3=\left\{\begin{pmatrix} 2 \\ -1 \\ -1 \end{pmatrix}, \begin{pmatrix} 2 \\ -\omega^2 \\ -\omega \end{pmatrix}, \begin{pmatrix} 2 \\ -\omega \\ -\omega^2 \end{pmatrix}, \begin{pmatrix} 0 \\ -\sqrt{3}\,i \\ \sqrt{3}\,i \end{pmatrix},\begin{pmatrix} 0 \\ -\sqrt{3}\,i\omega^2 \\ \sqrt{3}\,i\omega \end{pmatrix},\begin{pmatrix} 0 \\ -\sqrt{3}\,i\omega \\ \sqrt{3}\,i\omega^2 \end{pmatrix}\right\}\frac{v_{\varphi3}}{\sqrt{6}}\,.
\label{eq:solution2}
\end{eqnarray}
Although the solutions in each class seem different, they have no essentially physical differences. To keep the charged leptom mass matrix under $Z_3$ diagonal in this basis, we choose $(1,0,0)^Tv_{\varphi1}$ and $(1,1,1)^Tv_{\varphi2}/\sqrt{3}$ to characterise $\langle \varphi \rangle_1$ and $\langle \varphi \rangle_2$, respectively, as shown in the maintext.

\end{document}